\documentclass[aps,prb,superscriptaddress,showpacs,twocolumn,bibnotes]{revtex4}
\usepackage{graphicx}
\usepackage{amssymb}
\usepackage{amsmath}

\newcommand{\bk}{{\bf k}}
\newcommand{\bq}{{\bf q}}

\newcommand{\bG}{{\bf G}}
\newcommand{\bK}{{\bf K}}

\newcommand{\bx}{{\bf x}}
\newcommand{\by}{{\bf y}}
\newcommand{\br}{{\bf r}}
\newcommand{\bR}{{\bf R}}
\newcommand{\bJ}{{\bf J}}
\newcommand{\bz}{{\bf 0}}
\newcommand{\ba}{{\bf a}}
\newcommand{\bh}{{\bf h}}

\renewcommand{\Im}{{\mathop{\rm{Im}}\nolimits\,}}
\renewcommand{\Re}{{\mathop{\rm{Re}}\nolimits\,}}

\newcommand{\kB}{k_{\mathrm{B}}}

\newcommand{\nF}{n_{\mathrm{F}}}

\newcommand{\Ret}{{\mathrm{R}}}

\newcommand{\modified}[1]{{\relax #1}}

\begin{document}

\title{Effect of impurities in high-symmetry lattice positions on the local
density of states and conductivity of graphene}

\author{F. M. D. Pellegrino}
\affiliation{Dipartimento di Fisica e Astronomia, Universit\`a di Catania,\\
Via S. Sofia, 64, I-95123 Catania, Italy}
\affiliation{Scuola Superiore di Catania, Via S. Nullo, 5/i, I-95123 Catania,
Italy}
\affiliation{INFN, Sez. Catania, I-95123 Catania, Italy}
\author{G. G. N. Angilella}
\email[Corresponding author. E-mail: ]{giuseppe.angilella@ct.infn.it}
\affiliation{Dipartimento di Fisica e Astronomia, Universit\`a di Catania,\\
Via S. Sofia, 64, I-95123 Catania, Italy}
\affiliation{Scuola Superiore di Catania, Via S. Nullo, 5/i, I-95123 Catania,
Italy}
\affiliation{INFN, Sez. Catania, I-95123 Catania, Italy}
\affiliation{CNISM, UdR Catania, I-95123 Catania, Italy}
\author{R. Pucci}
\affiliation{Dipartimento di Fisica e Astronomia, Universit\`a di Catania,\\
Via S. Sofia, 64, I-95123 Catania, Italy}
\affiliation{CNISM, UdR Catania, I-95123 Catania, Italy}

\date{\today}

\begin{abstract}
\medskip

Motivated by quantum chemistry calculations, showing that molecular adsorption
in graphene takes place on preferential sites of the honeycomb lattice, we study
the effect of an isolated impurity on the local electronic properties of a
graphene monolayer, when the impurity is located on a site-like, bond-like, or
hollow-like position. We evaluate the local density of states (LDOS) as a
function of energy on the impurity and on its neighboring sites, as well as in
reciprocal space, at an energy corresponding to a bound state, in the three
cases of interest. The latter study may be relevant to interpret the results of
Fourier transformed scanning tunneling spectroscopy, as they show which states
mostly contribute to impurity-induced variations of the LDOS. We also estimate,
semi-analytically, the dependence of the condition for having a low-energy bound
state on the impurity potential strength and width. Such results are then
exploited to obtain the quasiparticle lifetime and the static conductivity in
graphene in the dilute impurity limit. In particular, we recover a sublinear
dependence of the conductivity on the carrier concentration as a generic
impurity effect.

\end{abstract} 

\maketitle

\section{Introduction}

Graphene is the two-dimensional allotrope of carbon, which is characterized by a
honeycomb lattice. Despite its structural simplicity, only recently it has been
obtained in laboratory \cite{Novoselov:04,Novoselov:05,CastroNeto:08}, thus
giving rise to a tremendous outburst of research activity, both among
experimentalists and theoreticians. Its remarkable electronic properties,
largely due to its reduced dimensionality, and its relatively high degree of
symmetry, make graphene an ideal candidate for applications in micro and
nanoelectronics. In particular, it has been recently suggested that charging can
be controlled at the atomic level, thereby enabling one to tailor some of the
magnetic properties of the system \cite{Uchoa:08}. On the other hand, its linear
quasiparticle dispersion relation suggests an analogy between the low-energy
excitations in graphene and relativistic massless particles, obeying Dirac-Weyl
equation, thus allowing the study of relativistic effects in a condensed matter
system \cite{Zhang:05,Berger:06}.

Since most of the intriguing physical properties of graphene stem from its
perfect crystal lattice, it is of interest to study how some of these are
affected by the presence of localized impurities. It is well-known that disorder
can significantly affect the electronic properties of graphene, especially when
the chemical potential traverses the Dirac points. This can be brought about not
only by impurities \cite{Pereira:06,Cheianov:06,Bena:07}, but also by
topological defects \cite{Vozmediano:05}, edges \cite{Peres:06}, substrate
corrugations \cite{Stolyarova:07}, and ripples \cite{Martin:08}.

Isolated short-range impurities have been shown to modify the local
single-particle electronic properties of graphene, such as the local density of
states (LDOS) \cite{Pereira:06,Peres:06,Wehling:07,Bena:07}, and can induce
Friedel oscillations \cite{Cheianov:06}. The role of strength, width and
concentration of impurities in altering the local energy spectrum has been
studied theoretically \cite{Skrypnyk:07,Peres:06}. \modified{The relevance of
special symmetries and how they manifest themselves in the scattering around
impurities has been emphasized in Ref.~\onlinecite{Basko:08}.} Moreover, the study of
the impurity effects on the LDOS is relevant to analyze the experimental results
of scanning tunneling microscopy (STM)
\cite{Ishigami:07,Stolyarova:07,Geringer:09}, and can elucidate the role of
correlations in the electron liquid in graphene. In particular, it has been
suggested that Fourier transformed scanning tunneling spectroscopy (FTSTS)
results can also be instrumental to identify experimentally monolayer and
bilayer graphene \cite{Bena:07,Mallet:07}.

Disorder is also known to affect considerably the transport properties of
graphene. In particular, the presence of disorder may explain the finite value
of the conductivity in pure graphene \cite{Geim:07,Hwang:07}. In the case of
graphene on a substrate, an inhomogeneous potential distribution may be brought
about by charged impurities located close to the substrate surface. At low
electron or hole concentration, this may induce sizeable spatial fluctuations of
the carrier concentration, and may therefore justify a nonzero conductivity,
even in the absence of any gate potential \cite{Hwang:07}. This has been
experimentally verified using a scanning single-electron transistor
\cite{Martin:08}. Such a regime of inhomogeneity persists beyond neutrality, and
characterizes also suspended graphene samples before annealing
\cite{Bolotin:08}. After annealing, the conductivity displays a sublinear
dependence on carrier concentration around zero doping, which may be due to
short-range impurities, such as point defects \cite{Tan:07,Stauber:08,Du:08}.

In this paper, we will be mainly concerned with the effects on the LDOS and on
the conductivity of graphene due to single or distributed impurities located in
a high-symmetry position of the honeycomb lattice. These include the sites of
the direct lattice, the position midway two neighboring carbon atoms, and the
center of the hexagon plaquettes. Such positions have been extensively studied
also within quantum chemical calculations, as they are expected to be favored in
the adsorption of hydrogen, water, and other simple molecules \cite{Forte:08a}.

After reviewing the formalism for a single localized impurity in graphene in
Sec.~\ref{sec:model}, we will present our results for the LDOS in the presence
of a single impurity, either in the site-like, bond-like, or hollow-like
configuration (Sec.~\ref{sec:single}). Our results include the energy dependence
of the LDOS on the impurity site and its nearest neighbors, and the reciprocal
lattice structure of the LDOS close to a resonance. Then, in
Sec.~\ref{sec:many}, we will generalize the above results in the case of many
impurities, in the dilute limit, within the full Born approximation. In
particular, we shall be interested in the case in which all impurities are
located in a preferential class of lattice sites. We will derive the LDOS in
reciprocal space in the case of many impurities, and discuss the dependence of
the quasiparticle lifetime on the impurity concentration. Finally, it will be
shown that, close to a low-energy resonance, disorder induces a sublinear
dependence of the conductivity on the carrier concentration, and that such an
effect is rather insensible to the impurity concentration, albeit in the dilute
limit. We summarize our results in Sec.~\ref{sec:conclusions}.

\section{Model}
\label{sec:model}

We begin by reviewing the \modified{tight-binding approximation and the}
$T$-matrix formalism for a single non-magnetic impurity in graphene
\cite{Bruus:04}. Within the tight-binding approximation, a graphene monolayer in
the presence of a single impurity localized at position $\bx$ will be described
by the Hamiltonian
\begin{equation}
H = \sum_{\bk\lambda} \xi_{\bk\lambda} c^\dag_{\bk\lambda} c_{\bk\lambda}
+ V_0 \Psi^\dag (\bx) \Psi(\bx) .
\end{equation}
Here, $c^\dag_{\bk\lambda}$ ($c_{\bk\lambda}$) is a creation (annihilation)
operator for a quasiparticle with wavevector $\bk$ within the first Brillouin zone
and band index $\lambda=1,2$, $\xi_{\bk\lambda} = E_{\bk\lambda}
-\mu$ is the tight-binding dispersion relation for band $\lambda$, measured with
respect to the chemical potential $\mu$, and $V_0$ is a measure of the strength of the
impurity potential. Expanding the field operators $\Psi^\dag(\bx)$, $\Psi(\bx)$
appearing in the impurity potential with respect to the
tight-binding basis states, one finds
\begin{equation}
H = \sum_{\bk\lambda} \xi_{\bk\lambda} c^\dag_{\bk\lambda} c_{\bk\lambda}
+ \sum_{\bk\bk^\prime\lambda\lambda^\prime} V_{\lambda\lambda^\prime}
(\bk,\bk^\prime ) c^\dag_{\bk\lambda} c_{\bk^\prime\lambda^\prime} ,
\end{equation}
where
\begin{equation}
V_{\lambda\lambda^\prime} (\bk,\bk^\prime ) = V_0 \psi^\ast_{\bk\lambda}
(\bx) \psi_{\bk^\prime \lambda^\prime} (\bx),
\label{eq:Vsep}
\end{equation}
and $\psi_{\bk\lambda} (\bx)$ is the Bloch wavefunction employed in the
tight-binding diagonalization of the pure sector of the Hamiltonian.


\subsection{Tight-binding approximation}

For the sake of completeness, we briefly review the main features of the tight
binding approximation employed in the present work. Graphene is characterized by
a honeycomb lattice, with basis vectors $\ba_1 = a(3,\sqrt{3})/2$ and $\ba_2 =
a(3,-\sqrt{3})/2$, where $a=0.142$~nm is the C--C distance \cite{CastroNeto:08}.
This is equivalent to two interpenetrating $A$ and $B$ triangular sublattices,
with nearest neighbor sites connected by the vectors $\delta_1 =
a(1,\sqrt{3})/2$, $\delta_2 = a(1,-\sqrt{3})/2$, $\delta_3 = a(-1,0)$.
Correspondingly, the first Brillouin zone in the reciprocal lattice is an
hexagon with vertices in the so-called Dirac points, $\bK = \frac{2\pi}{3a}
(1,\frac{\sqrt{3}}{3} )$, $\bK^\prime = \frac{2\pi}{3a}
(1,-\frac{\sqrt{3}}{3})$.

A suitable choice within the standard tight-binding approximation consists in
retaining hopping and overlap terms between nearest neighbor sites
\cite{Saito:98}. This gives rise to the two bands
\begin{equation}
E_{\bk\lambda} = \frac{\pm t |\gamma_\bk|}{1\mp s|\gamma_\bk|} ,
\label{eq:banddisp}
\end{equation}
where the bottom and top signs apply to the valence band, with $\lambda=1$,
and conduction band, with $\lambda=2$, respectively. In Eq.~(\ref{eq:banddisp}),
$t=2.8$~eV and $s=0.07$ are the nearest neighbor hopping and overlap parameters,
respectively \cite{Reich:02}, and 
\begin{equation}
\gamma_\bk = \sum_{\ell=1}^3 e^{i\bk\cdot\delta_\ell}
\label{eq:structfac}
\end{equation}
is the usual (complex) structure factor in momentum space. In the limit $s=0$,
one recovers the symmetry between valence and conduction bands, $E_{\bk\lambda}
= \pm t |\gamma_\bk|$.

One has still a choice to fix the functional form of the Bloch wavefunctions
that define the basis set implied in the tight binding approximation. These are
linear combination of tightly bound atomic functions, and will therefore be
termed pseudoatomic wavefunctions in the following. 
\modified{%
The approximation of using pseudoatomic wavefunctions with a finite extension,
while retaining a localized impurity potential, allows one to treat exactly also
the case in which a short-range impurity is located in an out-of-lattice
position, as is the case of hollow-like impurities addressed to below
(Sec.~\ref{ssec:hollow}).}
Due to the two-dimensionality of the graphene sheet, we can safely neglect their
extension along the axis orthogonal to the graphene plane, $z$ say.

One possible choice is such that its square modulus is a normalized gaussian
\cite{Bena:09}
\begin{equation}
\phi_g (\br) = \frac{1}{2\sqrt{3\pi}} \frac{Z_g}{a} \exp (-\rho_g^2 /24 ) ,
\label{eq:gaussian}
\end{equation}
where $\rho_g = Z_g r/a$, and $Z_g$ can be used to tune the spatial extension of
the wavefunction, characterized by an average radius $\bar{r}_g = \langle x^2 +
y^2 \rangle_g^{1/2} = 2\sqrt{3}a/Z_g$.

Another possible choice is such that its square modulus is a normalized combination of
modified Bessel functions of second kind \cite{GR}
\begin{equation}
\phi_b (\br) = \frac{1}{4\sqrt{\pi}} \frac{Z_b}{a} \sqrt{2\rho_b K_1(\rho_b ) +
\rho_b^2 K_0 (\rho_b )},
\label{eq:bessel}
\end{equation}
where $\rho_b = Z_b r/a$, and $Z_b$ is again a parameter related to the spatial
extension. Like the gaussian pseudoatomic wavefunction, Eq.~(\ref{eq:gaussian}),
also Eq.~(\ref{eq:bessel}) has a bell-shaped behavior, but decays more slowly,
$\phi_b (\br) \sim \rho_b^{3/4} \exp(-\rho_b)$, for $\rho_b\gg1$. The
expectation value of any cylindrically symmetric function with respect to
Eq.~(\ref{eq:bessel}) is the same as the expectation value with respect to the
$2p_z$ hydrogenic wavefunction with atomic number $Z_b$. In particular, the
average radius is similarly given by $\bar{r}_b = \langle x^2 + y^2
\rangle_b^{1/2} = 2\sqrt{3}a/Z_b$.

In both cases, the parameters are fixed by the condition that the nearest
neighbor overlap integral yields \cite{Reich:02} $s=0.07$, so that $Z_g = 11.2$
and $Z_b = 12.8$. In the case of a single impurity, we have numerically verified
that all results do not qualitatively depend on the particular choice of the
pseudoatomic wavefunction, and that any quantitative difference is within
graphical resolution. This is because the impurity effects considered here
depend mainly on the short-distance behavior of $\phi(\br)$. Therefore, in this
paper we have chosen to present results obtained with the gaussian choice for
the pseudoatomic wavefunctions, Eq.~(\ref{eq:gaussian}). In terms of these, the
Bloch wavefunction on which the tight-binding approximation is based is
\begin{equation}
\psi_{\bk\lambda}(\br) = \frac{1}{\sqrt{N}} \sum_j \phi(\br-\bR_j^\lambda )
e^{i\bk\cdot\bR_j^\lambda}, 
\label{eq:Blochband}
\end{equation}
where $\bR_j^\lambda$ are vectors of the $\lambda=A$ and $B$ sublattices,
respectively.

\modified{%
Let $\psi_{\bk\mu}$ ($\mu=A,B$) denote the Bloch wavefunctions in the sublattice
representation. These are then related to the Bloch wavefunctions
$\psi_{\bk\lambda}$ ($\lambda=1,2$) in the band representation,
Eq.~(\ref{eq:Blochband}), by the unitary transformation
\begin{equation}
\psi_{\bk\lambda} = \sum_{\mu=A,B} U_{\lambda\mu} (\bk) \psi_{\bk\mu} ,
\end{equation}
where $U_{\lambda\mu} (\bk)$ is the generic element of the matrix
\begin{equation}
U (\bk) = \frac{1}{\sqrt{2}} \begin{pmatrix} 1 & -1 \\
e^{-i\theta_\bk} & e^{-i\theta_\bk}
\end{pmatrix} ,
\end{equation}
and
\begin{equation}
e^{i\theta_\bk} = - \frac{\gamma_\bk}{|\gamma_\bk |} ,
\label{eq:expstructfac}
\end{equation}
with $\gamma_\bk$ defined in Eq.~(\ref{eq:structfac}).}

\subsection{$T$-matrix formalism}

We then introduce the finite-temperature Green's functions
\begin{equation}
{\mathcal{G}}_{\lambda\lambda^\prime} (\bk,\bk^\prime,\tau) = - \langle T_\tau
[c_{\bk\lambda} (\tau) c^\dag_{\bk^\prime\lambda^\prime} (0) ] \rangle ,
\end{equation}
where $\langle\cdots\rangle$ is a quantum statistical average with respect
to $H$ at temperature $T$, and $T_\tau$ denotes ordering with respect to the
imaginary time $\tau$. Making use of the fermionic Matsubara frequencies
$\hbar\omega_n = (2n+1)\pi \kB T$, where $\hbar$ is Planck's constant and $\kB$
is Boltzmann's constant, one finds the usual Dyson's equation
\begin{eqnarray}
{\mathcal{G}}_{\lambda\lambda^\prime} (\bk,\bk^\prime,i\omega_n ) &=&
\delta_{\lambda\lambda^\prime} \delta_{\bk\bk^\prime} 
\mathcal{G}^0_{\lambda} (\bk,i\omega_n ) \nonumber\\
&&
\hspace{-2.5truecm}
+
\sum_{\bq\lambda^{\prime\prime}}
{\mathcal{G}}_{\lambda\lambda^{\prime\prime}} (\bk,\bq,i\omega_n )
V_{\lambda^{\prime\prime} \lambda^\prime} (\bq,\bk^\prime )
\mathcal{G}^0_{\lambda^\prime} (\bk^\prime,i\omega_n ) ,
\label{eq:Dyson}
\end{eqnarray}
where $\mathcal{G}^0_\lambda (\bk,i\omega_n ) = (i\omega_n
-\xi_{\bk\lambda})^{-1}$ is the Green's function of the pure system. Dyson's
equation (\ref{eq:Dyson}) can be solved iteratively by exploiting the fact that
the impurity potential is factorizable in momentum space \modified{(see
Appendix~\ref{app:Tmatrix}).} One finds
\begin{eqnarray}
{\mathcal{G}}_{\lambda\lambda^\prime} (\bk,\bk^\prime,i\omega_n ) &=&
\delta_{\lambda\lambda^\prime} \delta_{\bk\bk^\prime} 
\mathcal{G}^0_{\lambda} (\bk,i\omega_n ) \nonumber\\
&&
\hspace{-2.5truecm}
+
\mathcal{G}^0_{\lambda} (\bk,i\omega_n ) 
T_{\lambda\lambda^\prime} (\bx; \bk,\bk^\prime ,i\omega_n )
\mathcal{G}^0_{\lambda^\prime} (\bk^\prime,i\omega_n ) ,
\label{eq:DysonT}
\end{eqnarray}
where
\begin{equation}
T_{\lambda\lambda^\prime} (\bx; \bk,\bk^\prime ,i\omega_n ) =
\frac{1}{N} \frac{V_0 \check{\psi}_{\bk\lambda}^\ast (\bx)
\check{\psi}_{\bk^\prime\lambda^\prime} (\bx)}{1-V_0 \mathcal{G}^0
(\bx,\bx,i\omega_n )}
\label{eq:Tmatrix}
\end{equation}
is the generic element of the $T$-matrix, $\check{\psi}_{\bk\lambda} (\bx) =
\sqrt{N} \psi_{\bk\lambda} (\bx)$ is a rescaled basis function, and
\begin{equation}
\mathcal{G}^0 (\br,\br^\prime,i\omega_n ) = \frac{1}{N} 
\sum_{\bq\lambda^\prime} 
\check{\psi}_{\bq\lambda^\prime} (\br) 
\mathcal{G}^0_{\lambda^\prime} (\bq,i\omega_n )
\check{\psi}^\ast_{\bq\lambda^\prime} (\br^\prime ) .
\label{eq:G0}
\end{equation}
Eq.~(\ref{eq:DysonT}) shows that the correction due to a single localized impurity
vanishes as $1/N$ in the thermodynamic limit, $N\to\infty$. Going back to real
space by means of Eq.~(\ref{eq:G0}), one finds the imaginary-time Green's
function at position $\br$
\begin{equation}
\mathcal{G}(\br,\br,i\omega_n ) = \mathcal{G}^0 (\br,\br,i\omega_n )
+ \frac{V_0 \mathcal{G}^0 (\br,\bx,i\omega_n ) \mathcal{G}^0 (\bx,\br,i\omega_n
)}{1-V_0 \mathcal{G}^0 (\bx,\bx,i\omega_n )} .
\label{eq:G0x}
\end{equation}
In what follows, we shall be interested in the local density of states (LDOS)
$\rho(\br,\omega)$, which is experimentally accessible through STM measurements
\cite{Ishigami:07,Stolyarova:07,Geringer:09}, and is related to the imaginary
part of the analytically continued Green's function through
\begin{equation}
\rho(\br,\omega) = -\frac{1}{\pi} \Im G(\br,\br,\omega) ,
\label{eq:LDOS}
\end{equation}
where $G(\br,\br,\omega) = \mathcal{G} (\br,\br,i\omega_n \to \omega + i0^+ )$.
It is straightforward to observe that the chemical potential enters $G(\omega)$
only as an additive constant to $\omega$. Therefore, we can set hereafter
$\mu=0$, thereby neglecting any contribution arising from chemical or electrical
doping, \emph{e.g.} through a gate voltage.

\begin{figure}[t]
\centering
\includegraphics[height=0.8\columnwidth,angle=-90]{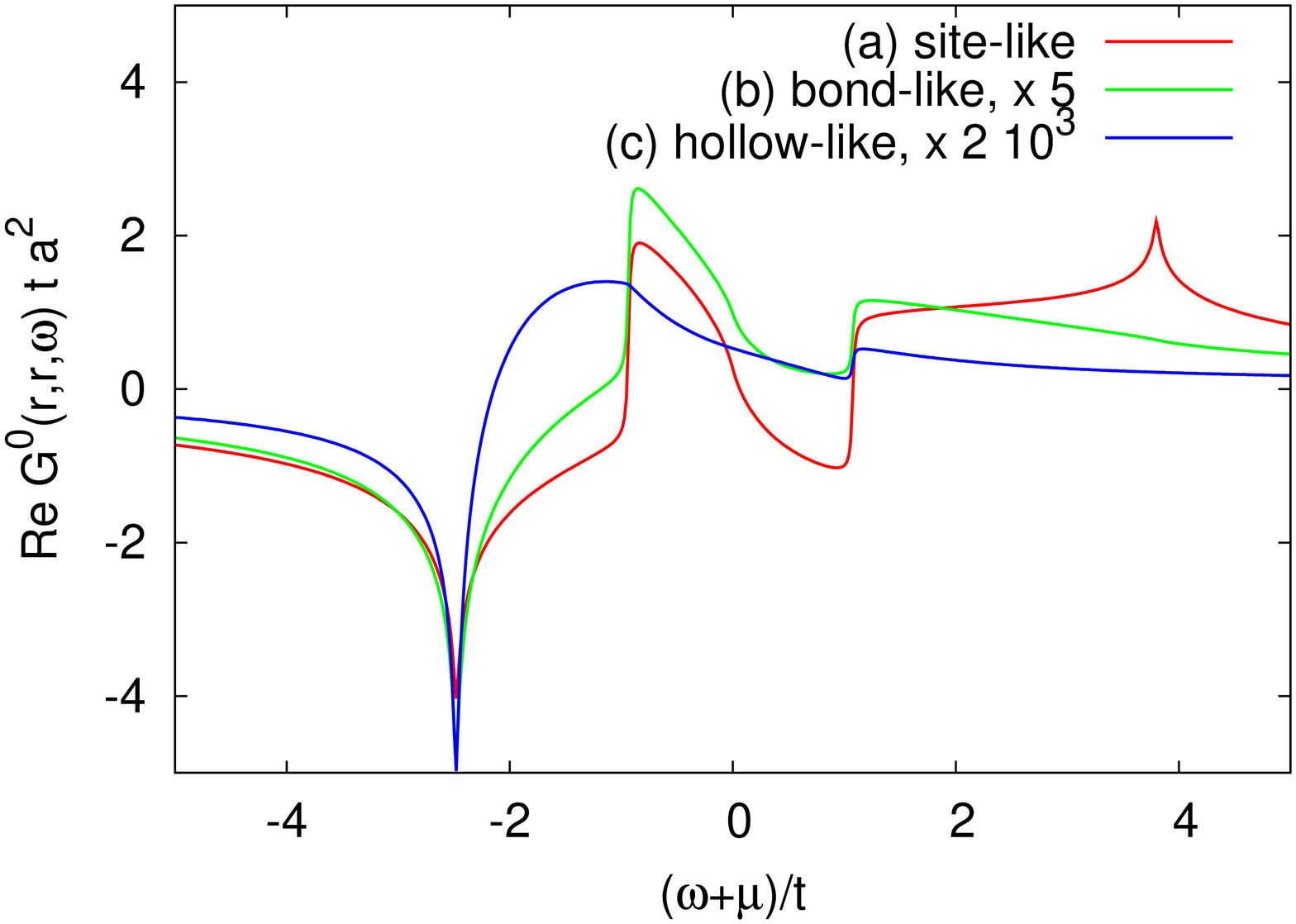}
\includegraphics[height=0.8\columnwidth,angle=-90]{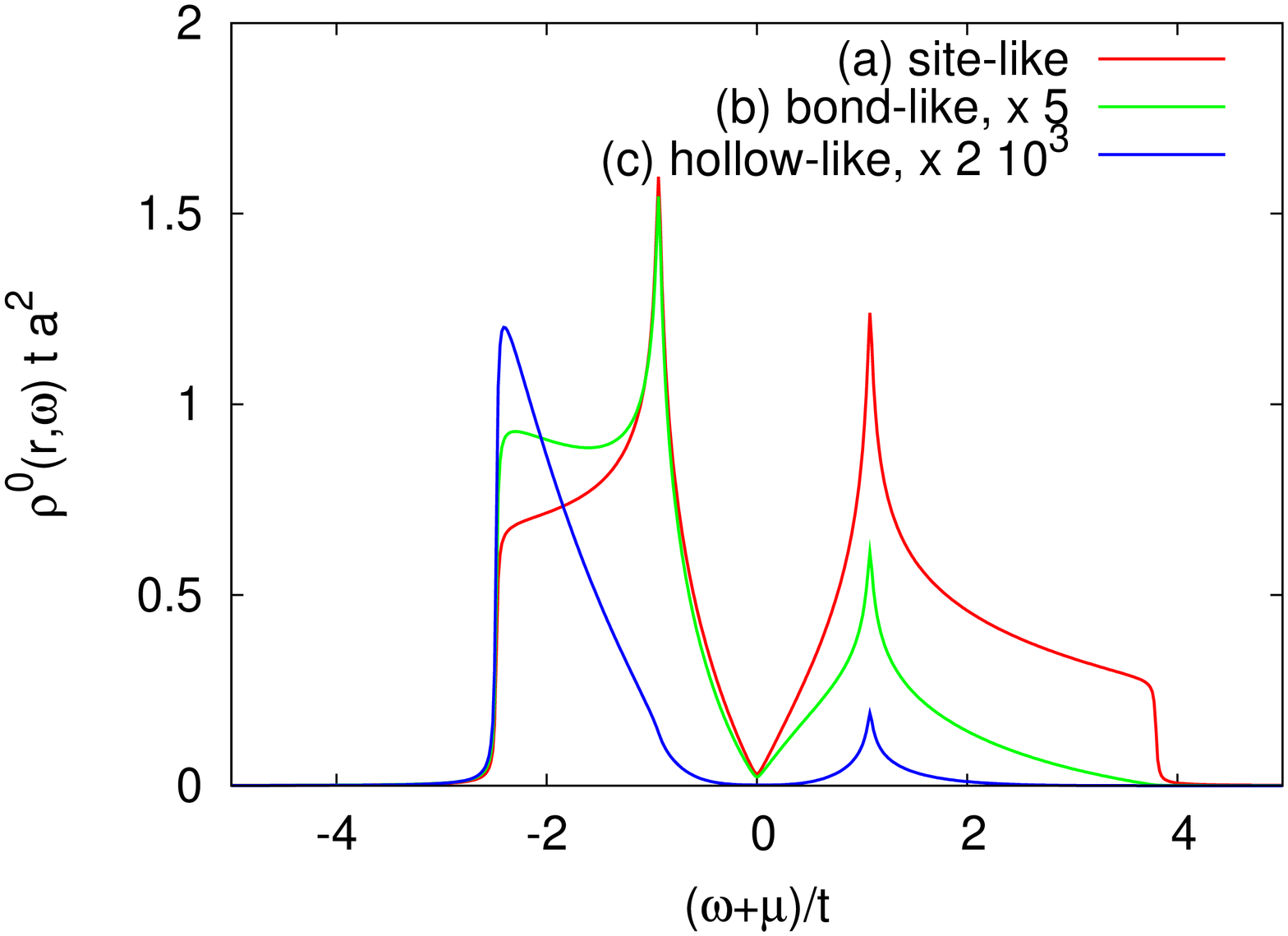}
\caption{(Color online) Real part of the unperturbed Green's function, $\Re G^0
(\bx,\omega)$ (top panel), and unperturbed LDOS, $\rho^0 (\bx,\omega) =
-\pi^{-1} \Im G^0 (\bx,\omega)$ (bottom panel), for the three cases of interest:
(a) site-like impurity ($\bx=\bz$), (b) bond-like impurity ($\bx=\delta_3/2$),
(c) hollow-like impurity ($\bx=-\delta_3$). It should be noticed that the latter
two cases have been scaled by the factors indicated in the caption.}
\label{fig:G0}
\end{figure}

Inspection of Eqs.~(\ref{eq:G0x}) and (\ref{eq:LDOS}) shows that the LDOS on the
impurity position ($\br=\bx$) is given by
\begin{equation}
\rho(\bx,\omega) = \frac{\rho^0 (\bx,\omega)}{[1-V_0 \Re G^0 (\bx,\bx,\omega)]^2
+ [\pi V_0 \rho^0 (\bx,\omega)]^2} ,
\label{eq:LDOSx}
\end{equation}
with $\rho^0 (\bx,\omega)$ denoting the LDOS at position $\br=\bx$ in the pure
case. In the limit of a vanishing unperturbed LDOS, $\rho^0 (\bx,\omega)\to0$,
one has $\rho(\bx,\omega) \to V_0^{-1} \delta [ 1 - V_0 \Re G^0
(\bx,\bx,\omega)]$. Such a circumstance is \emph{e.g.} realized below the
valence band, $\omega\leq\omega_\bot \approx -2.48 t$, or above the conduction
band, $\omega\geq\omega_\top\approx 3.80t$.  Direct inspection of $\Re G^0
(\bx,\bx,\omega)$ as a function of $\omega$ (Fig.~\ref{fig:G0}) leads to the
existence of a bound state outside the two bands, for a wide range of potential
strengths $V_0$. In particular, a bound state at $\omega<0$ may be formed below
the valence band only for some $V_0 < 0$, or above the conduction band for $V_0
> 0$, the energy of such a bound state moving farther from the bands, as $|V_0
|$ increases.

For future reference, it is also of interest to quote the expression of the LDOS
close to a bound state in reciprocal space, which reads
\begin{eqnarray}
\rho_\lambda (\bk,\omega) &=& -\frac{1}{\pi} \Im G_\lambda (\bk,\omega)
\nonumber\\
&=&
\frac{V_0}{N}
\frac{|\check{\psi}_{\bk\lambda} (\bx)|^2}{|\omega -\xi_{\lambda\bk}|^2}
\delta [ 1 - V_0 \Re G^0 (\bx,\bx,\omega)].
\label{eq:LDOSk}
\end{eqnarray}
\modified{%
In other words, Eq.~(\ref{eq:LDOSk}) applies to states with a vanishing
unperturbed LDOS, \emph{i.e.} to frequencies $\omega$ such that $\rho^0
(\bk,\omega)=0$ for all wavevectors $\bk$ in the 1BZ. This corresponds to
$\omega<\omega_\bot$, $\omega=0$, and $\omega>\omega_\top$.}

\section{Single impurity}
\label{sec:single}

In what follows, we shall analyze the effect on the LDOS, Eq.~(\ref{eq:LDOS}),
due to a single impurity localized in several high-symmetry positions of the
primitive cell in the graphene honeycomb lattice. These include an $A$ or $B$
site, usually occupied by a carbon atom (site-like impurity), the position
midway between an $A$ and $B$ site (bond-like impurity), and the position at the
center of an hexagon plaquette (hollow-like impurity).

\subsection{Site-like impurities}

Let us start by considering an impurity located on an $A$ or $B$ site
(say $\bx=\bz$, for definiteness). Such an impurity preserves the $D_{3h}$ symmetry.
This could be used to model a hydrogen impurity adsorbed on a carbon atom
\cite{Basko:08,Forte:08a}, or a vacancy (here obtained in the $V_0\to\infty$
limit) \cite{Pereira:08,Forte:08a}, as could be induced by proton irradiation
\cite{Esquinazi:03}.

\begin{figure}[t]
\centering
\includegraphics[height=0.8\columnwidth,angle=-90]{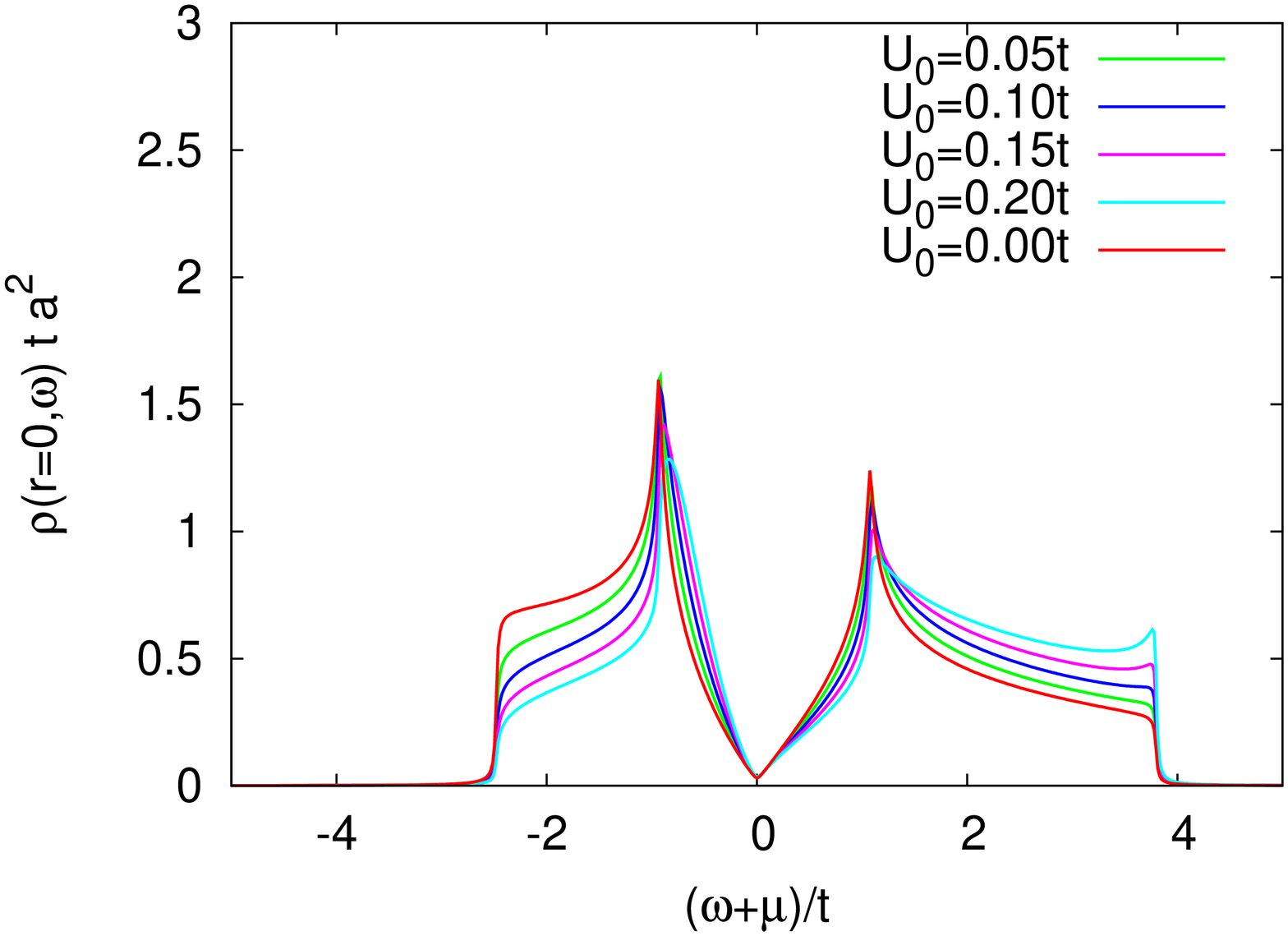}
\includegraphics[height=0.8\columnwidth,angle=-90]{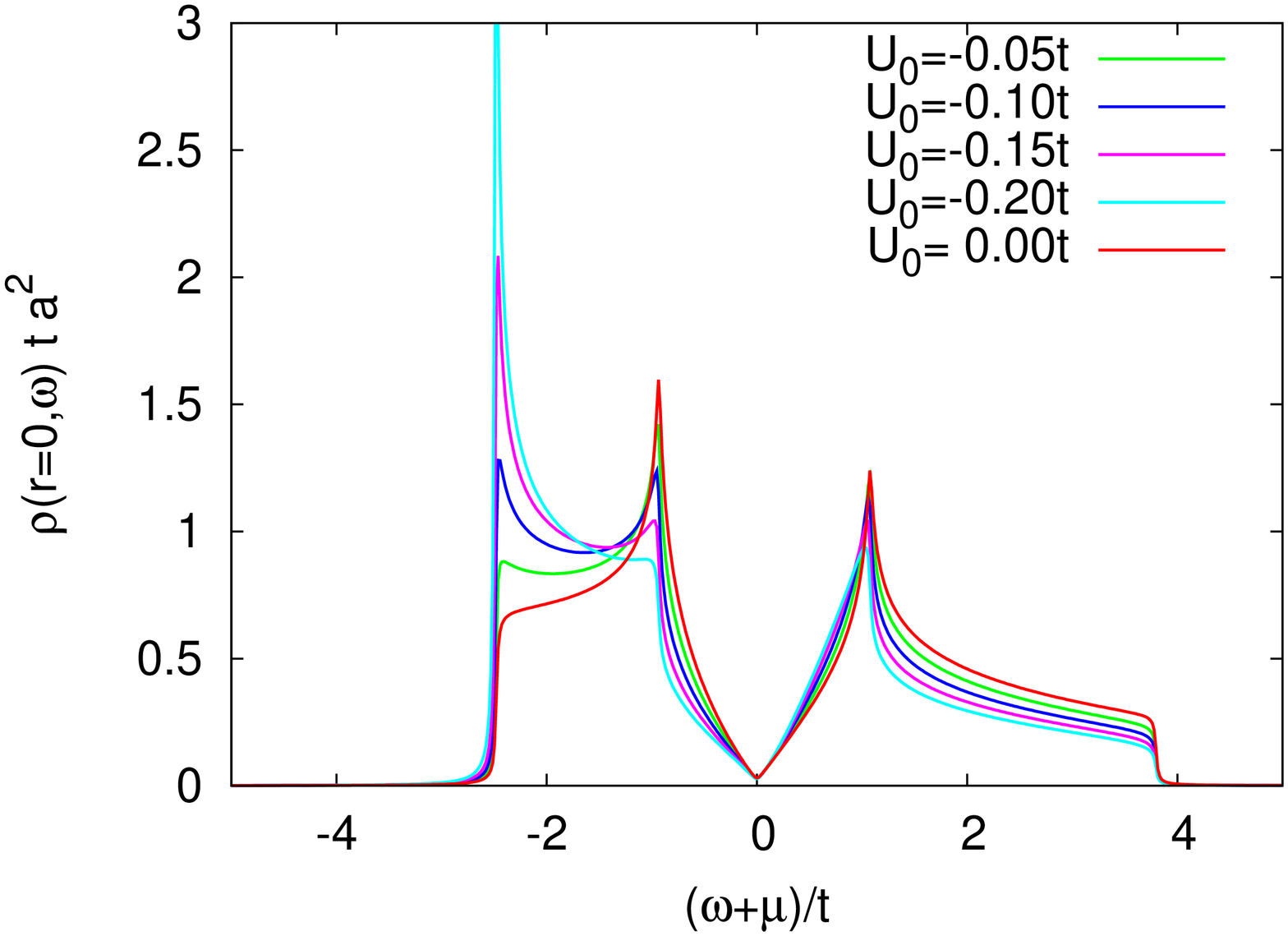}
\caption{(Color online) Local density of states $\rho(\bx=\bz,\omega)$,
Eq.~(\ref{eq:LDOSx}), on a site-like impurity located at $\bx=\bz$. Top panel
shows the LDOS for $U_0 /t = -0.05$, $-0.10$, $-0.15$, $-0.20$. Bottom panel
shows the LDOS for $U_0 /t = 0.05$, $0.10$, $0.15$, $0.20$. In both panels, we
also show the LDOS in the unperturbed case ($U_0 /t =0$).}
\label{fig:rhor_ona}
\end{figure}

Fig.~\ref{fig:rhor_ona} shows the LDOS on a single site-like impurity,
Eq.~(\ref{eq:LDOSx}), for negative as well as for positive values of the
impurity strength $U_0 = V_0 /a^2$, where $a$ is the lattice step, in the limit
of weak scattering ($|U_0|\ll6t$). As anticipated, a bound state forms below the
valence band if $U_0 <0$, whereas a bound state forms above the conduction band
for $U_0>0$.

From Eq.~(\ref{eq:LDOSx}), a resonance is formed at an energy $\omega_{res}$
between the two Van~Hove singularities when
\begin{equation}
1-V_0 \Re G^0 (\bx,\bx,\omega_{res})=0.
\label{eq:rescond}
\end{equation}
By inspection of the $\omega$-dependence of $\Re G^0 (\omega)$
(Fig.~\ref{fig:G0}), it follows that Eq.~(\ref{eq:rescond}) is fulfilled for
$-1 < t/U_0 < \frac{1}{2}$. Such a resonance is better resolved
when the unperturbed LDOS $\rho^0 (\bx,\bx,\omega)$ is weak for
$\omega\approx\omega_{res}$. This is indeed the case in the proximity of the
Dirac points, where $\rho^0 (\bx,\bx,\omega) \to0$ linearly as $\omega\to0$. A
special case is represented by the limit $U_0\to\infty$, corresponding to a
vacancy formation. In this case, the condition for a resonance is fulfilled at
$\omega\approx st$, where however the LDOS is strongly depressed.

\begin{figure}[t]
\begin{center}
\begin{minipage}[c]{0.48\columnwidth}
\begin{center}
\includegraphics[height=\textwidth,angle=-90]{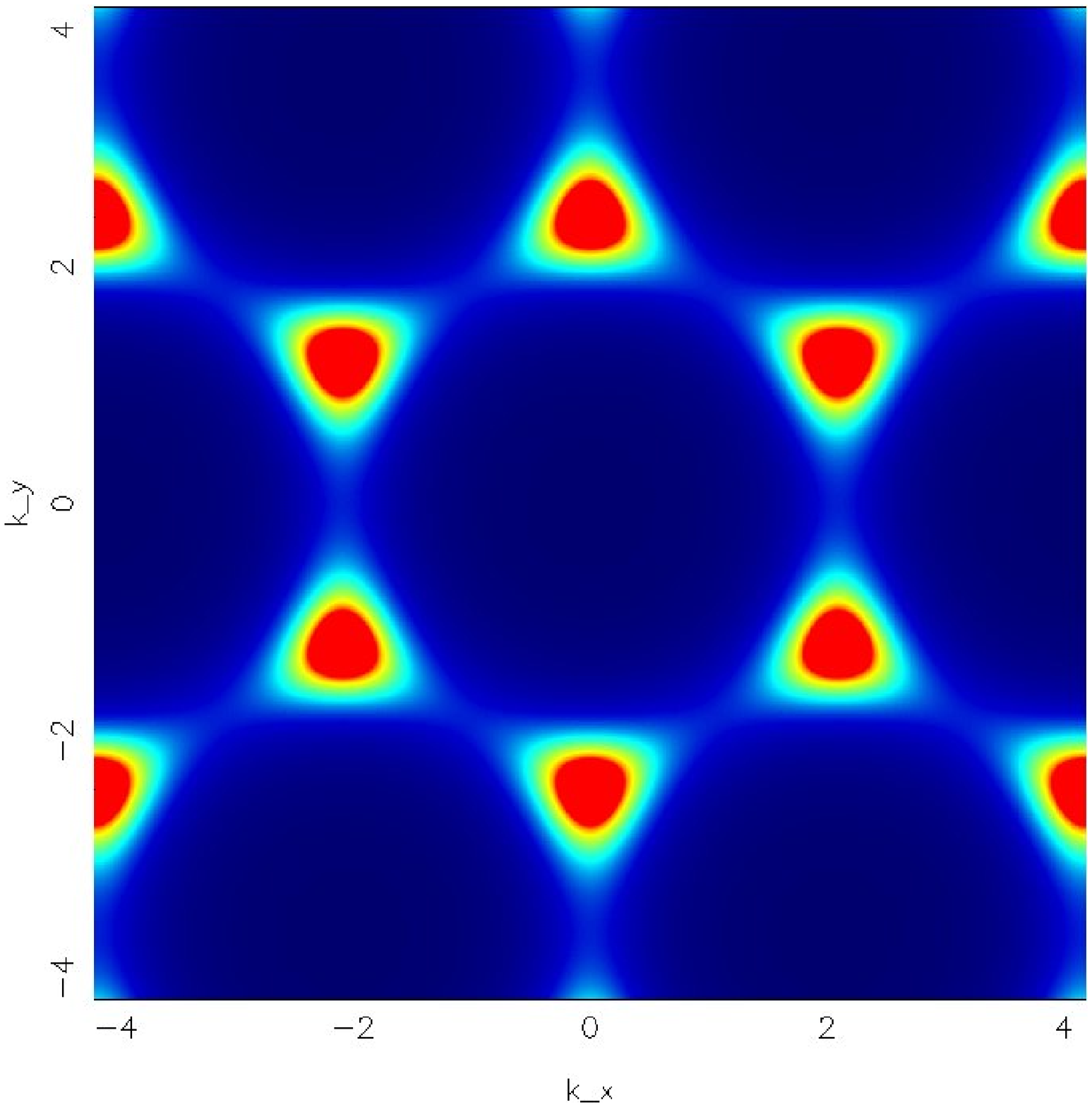}
\end{center}
\end{minipage}
\begin{minipage}[c]{0.48\columnwidth}
\begin{center}
\includegraphics[height=\textwidth,angle=-90]{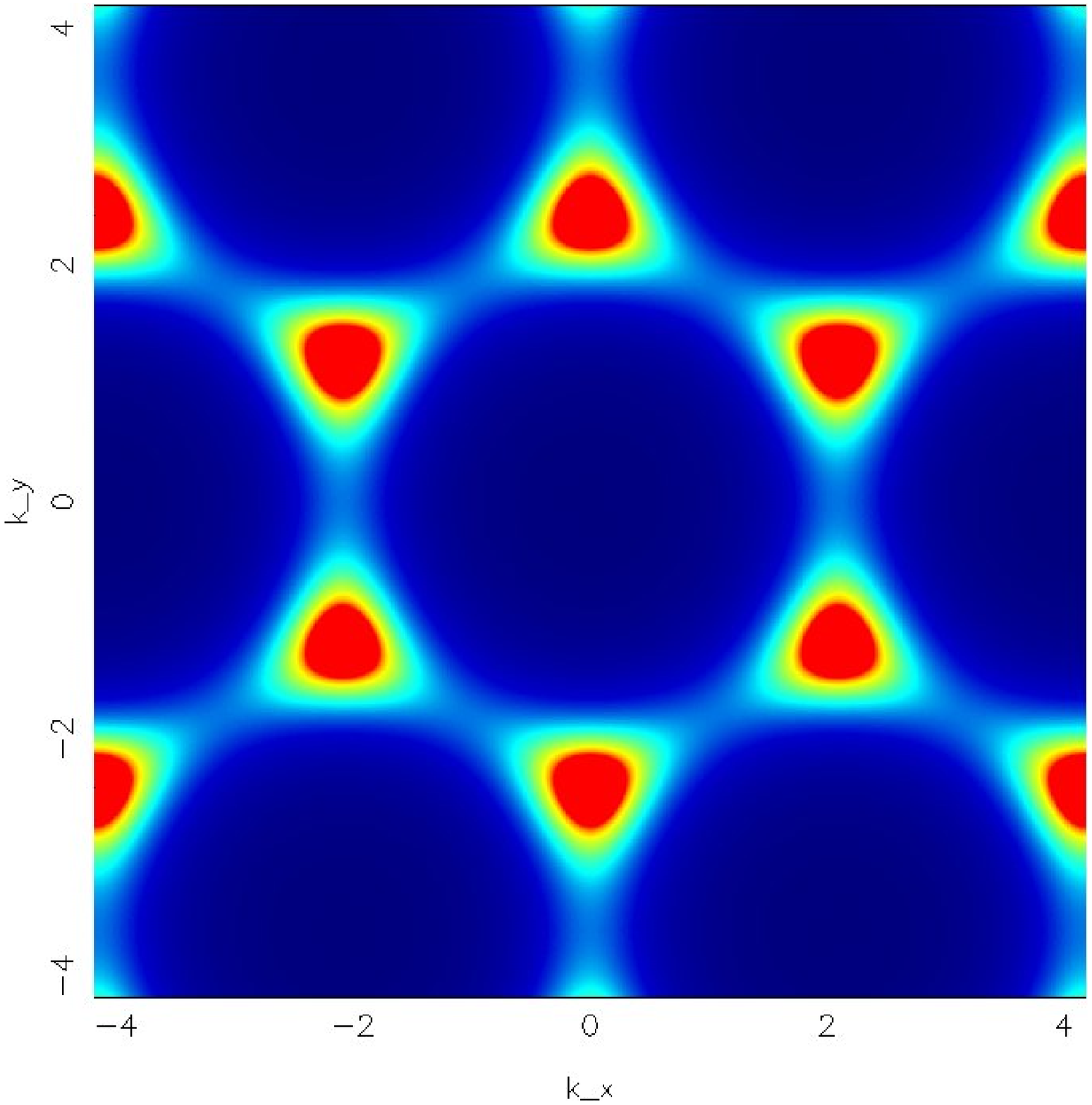}
\end{center}
\end{minipage}
\caption{(Color online) Contour plots of the LDOS in momentum space,
$\rho_\lambda (\bk,\omega=0)$,  Eq.~(\ref{eq:LDOSk}), for the valence
($\lambda=1$, left panel) and conduction band ($\lambda=2$, right panel). Here,
we are considering a site-like impurity with $U_0 = 3.7t$, thus giving rise to a
bound state at $\omega=0$.}
\label{fig:rhow0onsite}
\end{center}
\end{figure}

At exactly $\omega=0$, \emph{e.g.} when the chemical potential traverses the
Dirac points, the resonance becomes a true bound state, since $\rho^0
(\bx,\bx,\omega)=0$. The value of the impurity potential $U_0$ allowing a bound
state at $\omega=0$ can be obtained within a semi-analytical approach (see
Appendix~\ref{app:tightbinding} for details). This is based on an expansion of $\Re
G^0 (\bx,\bx,\omega=0)$ in Eq.~(\ref{eq:rescond}) at $\bx=\bz$. To the leading
terms in the nearest neighbors, one finds that the condition for having a bound
state at $\omega=0$ is
\begin{equation}
\frac{t}{U_0} \approx \phi (\bz) [s \phi (\bz) + 2 \phi(\delta_1 )],
\label{eq:rescond1}
\end{equation}
where $\phi(\bx)$ is a gaussian pseudo-atomic wavefunction, $s$ is the band
asymmetry parameter, and $\delta_1$ the position of a nearest neighbor to
$\bx=\bz$ (Appendix~\ref{app:tightbinding}). The first contribution to
Eq.~(\ref{eq:rescond1}) is due to the band asymmetry ($s\neq0$), whereas the second
contribution is related to the wavefunction width, and can be neglected for a
sufficiently localized pseudo-atomic wavefunction. In the limit of symmetric
bands ($s=0$) and localized wavefunctions, one recovers a bound state at exactly
$\omega=0$ in the case of a vacancy ($U_0 = \infty$) \cite{Pereira:06,Peres:06,Skrypnyk:07}.
\modified{%
From Eq.~(\ref{eq:rescond1}) one may conclude that a bound state is formed also
in the case of a localized pseudoatomic wavefunction, provided one retains a
nonzero band asymmetry ($s\neq0$), and that this takes place for a finite value
of the impurity potential ($U_0 < \infty$), in agreement with the findings of
Ref.~\onlinecite{Pereira:06}.}
Fig.~\ref{fig:rhow0onsite} shows a contour plot of the LDOS in momentum space
for an impurity potential generating a bound state at $\omega=0$, for both the
valence and conduction bands. In both cases, the largest contribution to
$\rho_\lambda (\bk,\omega)$ comes from the wavevectors close to the Dirac points.
Slight differences between the two bands are due to a nonzero asymmetry
parameter $s$.

\begin{figure}[t]
\centering
\includegraphics[height=0.8\columnwidth,angle=-90]{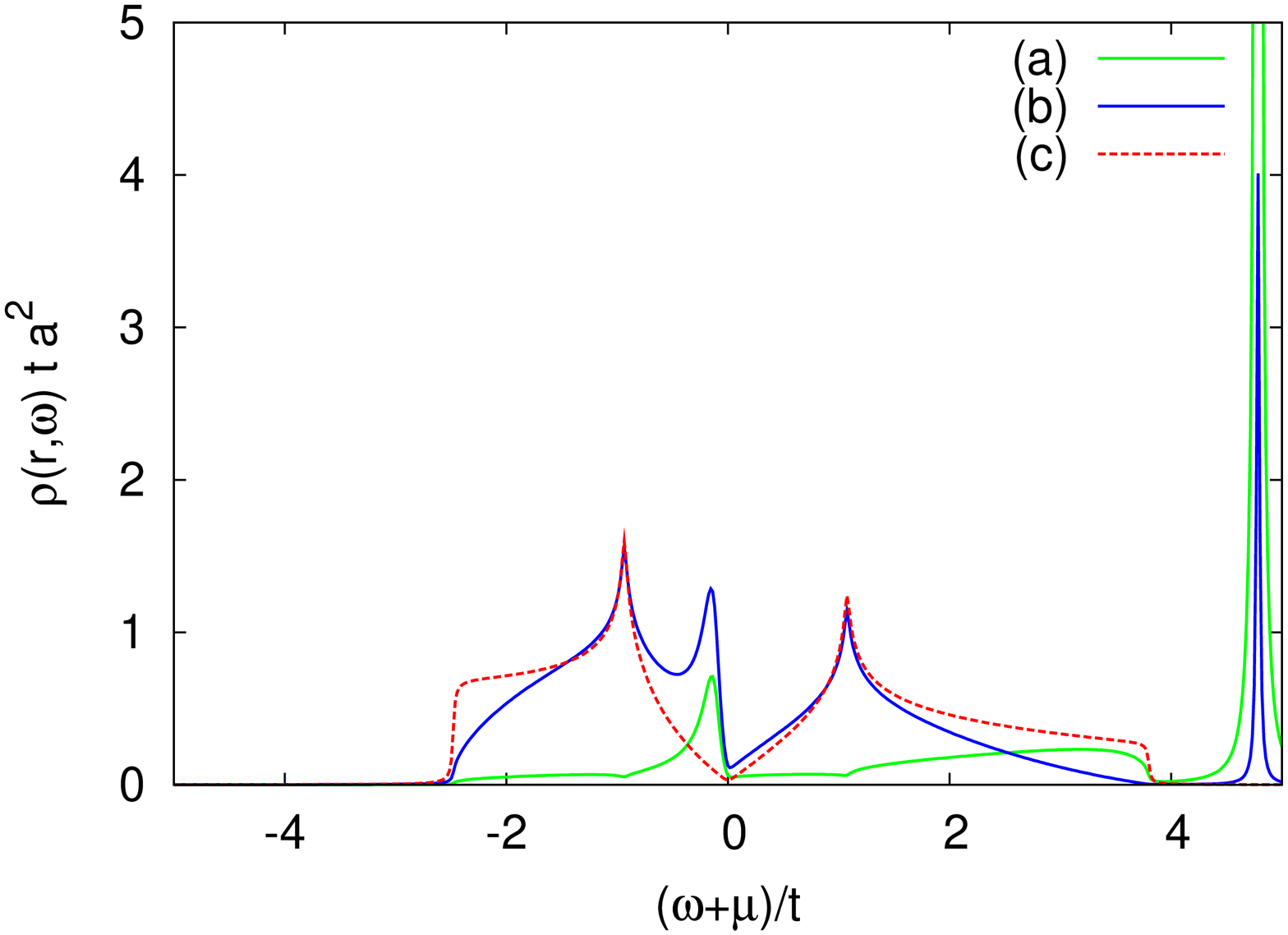}
\includegraphics[height=0.8\columnwidth,angle=-90]{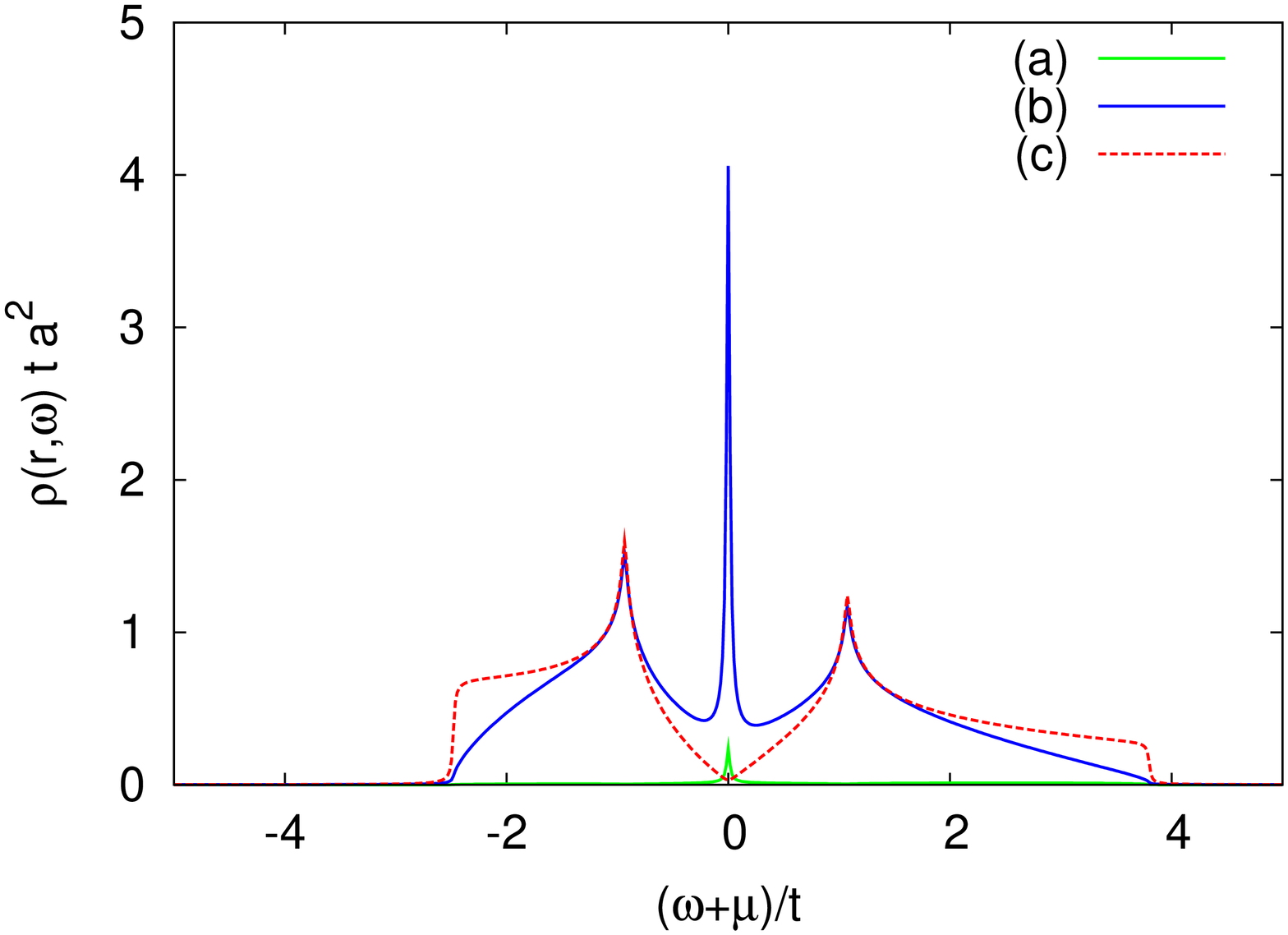}
\caption{(Color online) Showing the LDOS for a site-like impurity at $\bx=\bz$
(a) on the impurity site, (b) on a nearest neighbor site, (c) on a generic
lattice site in the unperturbed case. Top panel refers to low potential strength
($U_0 = 1.1t$), while bottom panel refers to large potential strength ($U_0 =
3.7t$).}
\label{fig:rhor_onatob}
\end{figure}

We end this subsection by considering the effect of a site-like impurity located
at $\bx=\bz$ on the LDOS at a neighboring site, $\by=\delta_3$, say. After the
appropriate analytical continuation, Eq.~(\ref{eq:G0x}) then yields
\begin{eqnarray}
\rho(\by,\omega) &\approx&  \frac{\Re G^0 (\by,\bx,\omega) \Re G^0
(\bx,\by,\omega)}{[\Re G^0 (\bx,\bx,\omega)]^2} \nonumber\\
&&\times V_0^{-1}\delta[1 - V_0 \Re G^0 (\bx,\bx,\omega)] ,
\label{eq:LDOSnn}
\end{eqnarray}
in the limit of vanishing unperturbed LDOS on the impurity, $\rho^0
(\bx,\omega)\to0$. Therefore, while the condition for having a bound state on a
neighboring site is the same as Eq.~(\ref{eq:rescond}), the relative weight with
respect to the impurity site is given by the prefactor in Eq.~(\ref{eq:LDOSnn}).
Fig.~\ref{fig:rhor_onatob} shows the LDOS on a neighboring site
($\by=\delta_3$), when a site-like impurity is located at $\bx=\bz$. The LDOS
corresponding to a bound state outside the band is larger on the impurity site
than on the neighboring site. The opposite is true for the LDOS corresponding to
the resonant state between the two Van~Hove singularities, which is depressed on
the impurity site than on the neighboring site. The same effect applies to the
bound state between the two Van~Hove singularities.

Such a behavior for resonant states in the energy range between the two Van~Hove
singularities is analogous to the one encountered in the $d$-density-wave (DDW)
phase, which has been suggested as a viable description of the pseudogap phase
in the high-$T_c$ cuprates \cite{Marston:89,Chakravarty:01}. In the DDW phase,
the LDOS on an atomic site vanishes linearly as $\omega\to0$ for the pure
system, and exhibits two Van~Hove singularities, symmetric with respect to
$\mu=0$. It has been demonstrated \cite{Zhu:01a,Morr:02} that a sufficiently
strong localized impurity produces a resonance between the two Van~Hove
singularities. The LDOS associated to such a resonance in the DDW phase is
larger on a nearest neighbor, than on the impurity site, in close analogy to
what is here shown for an impurity in graphene, \modified{and in agreement with
the findings of Ref.~\onlinecite{Wehling:07}.}
In both cases, the quasiparticle
bands are characterized by two inequivalent minima (the two Dirac points, in the
case of graphene), so that scattering processes due to short-range impurities
can be classified as intra-valley or inter-valley, depending on whether initial
and final states lie close to the same or to different extrema, respectively. The
peculiar behavior of the LDOS corresponding to resonant states at
$\omega\approx0$ is related to inter-valley scattering, at variance with bound
states outside the bandwidth. It is also relevant to note, in this context, that
single-impurity scattering around a localized impurity has been suggested as a
tool to distinguish between a DDW and a pseudogap phases, within the precursor
pairing scenario \cite{Andrenacci:04c}.

\subsection{Bond-like impurities}

\begin{figure}[t]
\centering
\includegraphics[height=0.45\columnwidth,angle=-90]{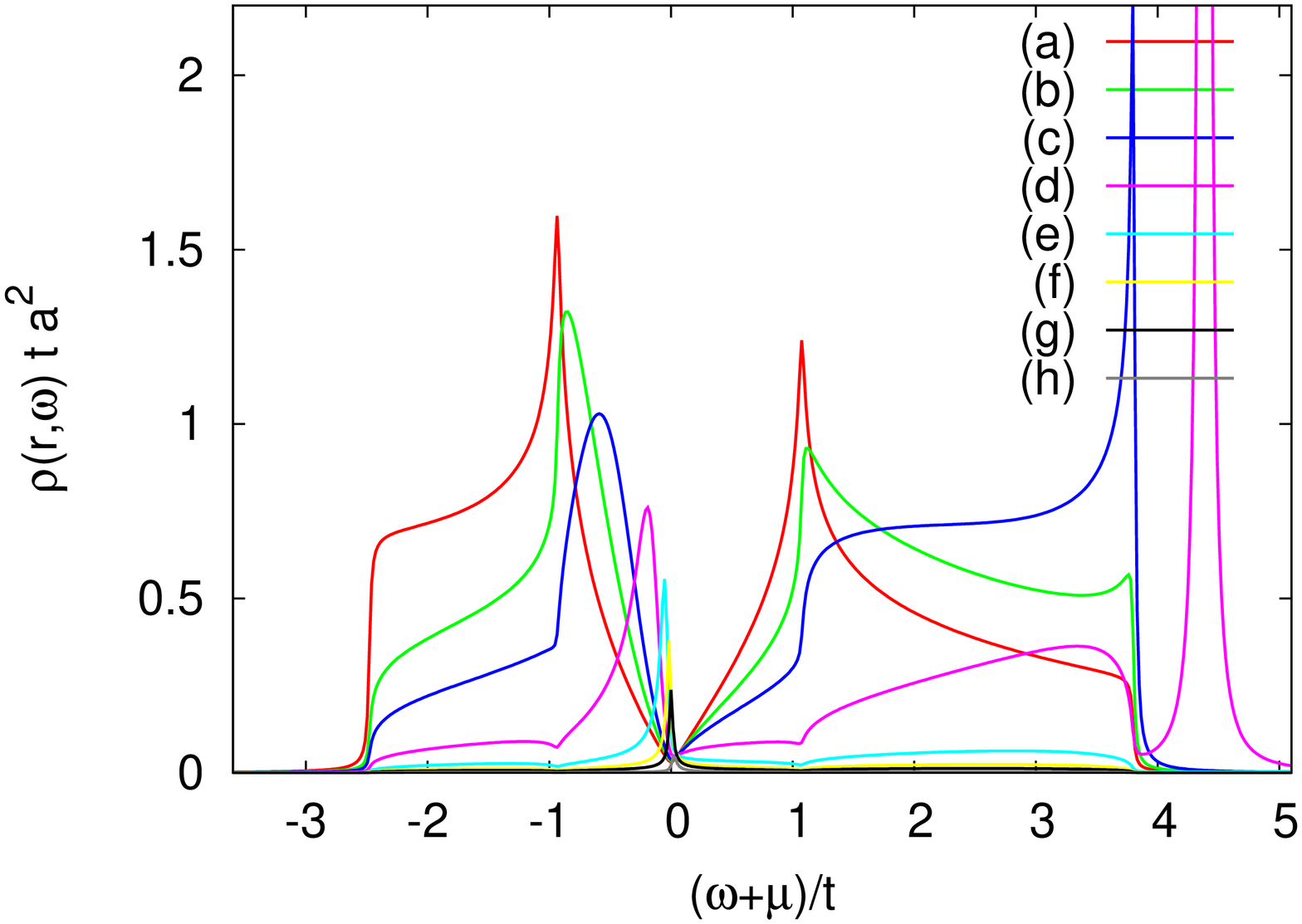}
\includegraphics[height=0.45\columnwidth,angle=-90]{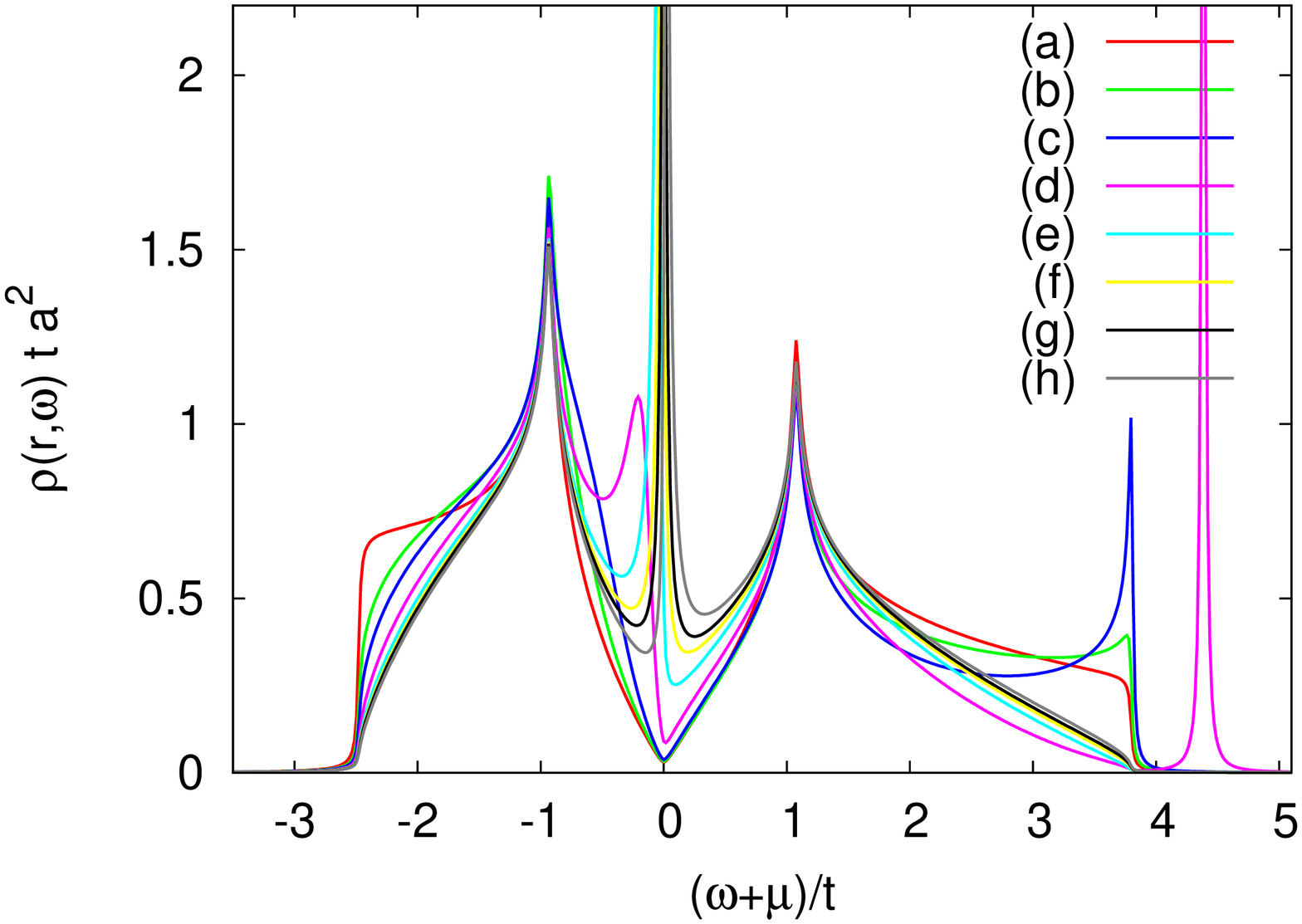}\\
\includegraphics[height=0.45\columnwidth,angle=-90]{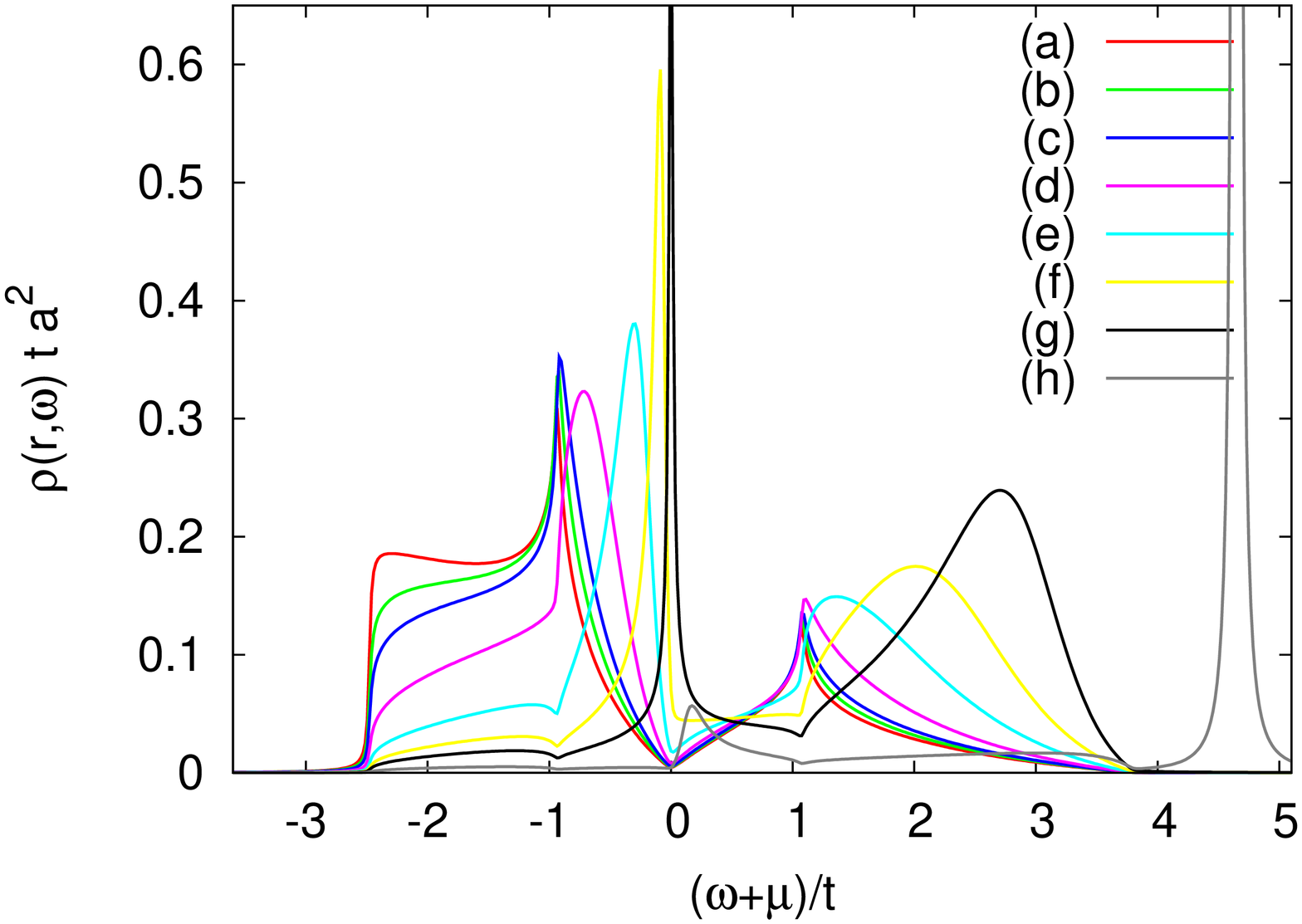}
\includegraphics[height=0.45\columnwidth,angle=-90]{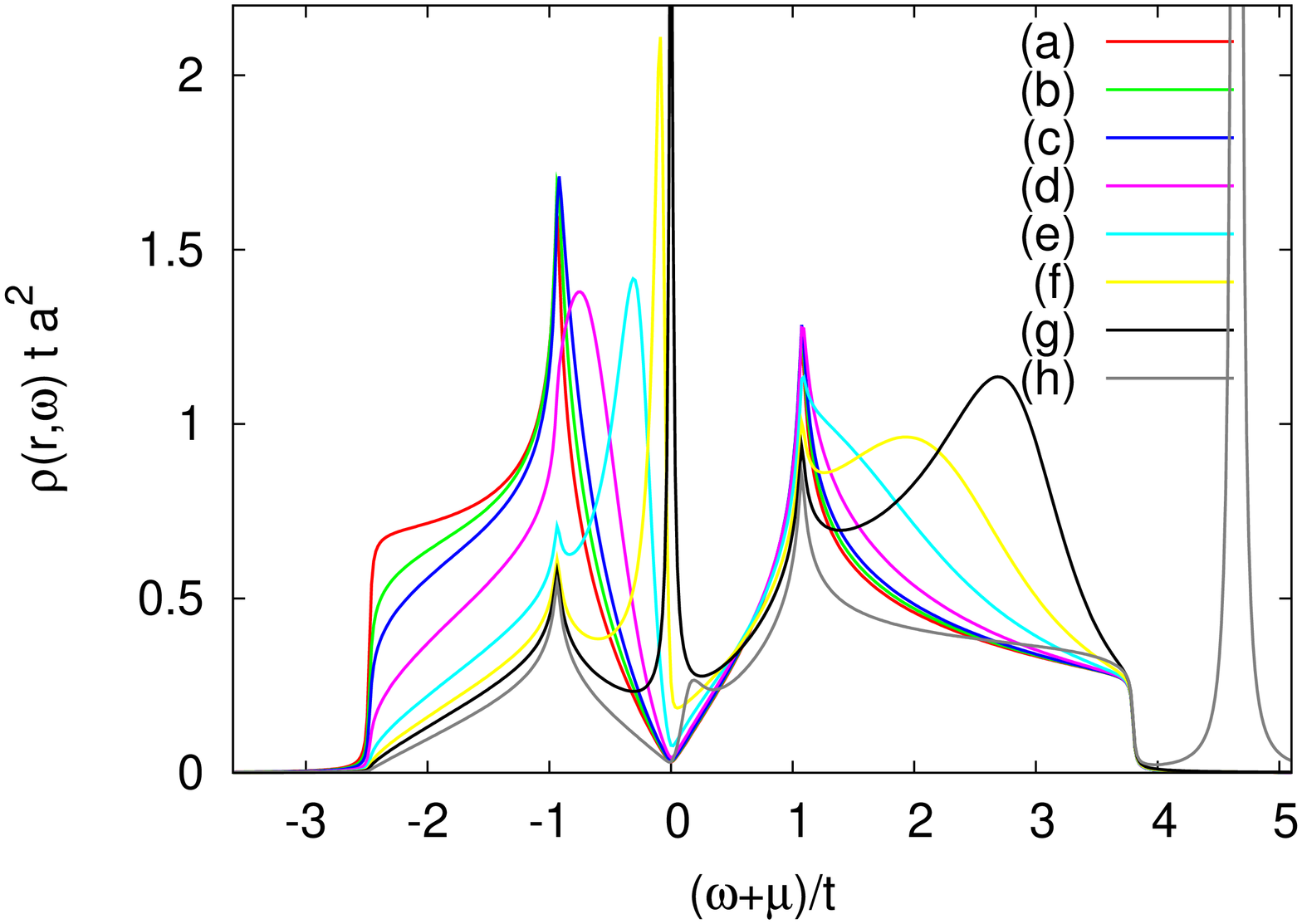}
\caption{(Color online) Local density of states for a site-like impurity (top
row panels), and a bond-like impurity (bottom row panels). Left panel on top refers to the
LDOS evaluated on the same position of a site-like impurity ($\br=\bx=\bz$);
right panel on top refers to the LDOS for a site-like impurity ($\bx=\bz$),
evaluated on a nearest neighbor lattice site ($\br=\delta_1$). Left panel on
bottom refers to the LDOS evaluated on the same position of a bond-like impurity
($\br=\bx=\delta_3/2$); right panel on bottom refers to the LDOS for a bond-like
impurity ($\bx=\delta_3/2$), evaluated on a nearest neighbor lattice site
($\br=\bz$).
The potential strengths are 
(a) $U_0 =0$; 
(b) $U_0 =0.05 \tilde{U}_0$; 
(c) $U_0 =0.10 \tilde{U}_0$;
(d) $U_0 =0.25 \tilde{U}_0$;
(e) $U_0 =0.50 \tilde{U}_0$;
(f) $U_0 =0.75 \tilde{U}_0$;
(g) $U_0 =\tilde{U}_0$;
(h) $U_0 =2.00 \tilde{U}_0$,
where $\tilde{U}_0$ is the value of $U_0$ yielding a bound state at $\omega=0$.}
\label{fig:bondtosite}
\end{figure}

An impurity located between an $A$ and $B$ site only preserves the $C_{2v}$
symmetry, and may be used to model an oxygen impurity between two carbon atoms
\cite{Basko:08}. This corresponds to three inequivalent positions in the real
lattice, although local effects on each of them are related by rotations of
multiples of $2\pi/3$. In the following, for definiteness, we shall therefore be
concerned with a bond-like impurity located at $\bx = \delta_3/2$.

With reference again to Eq.~(\ref{eq:LDOSx}), one finds a markedly different
$\omega$-dependence of $G^0 (\bx,\bx,\omega)$ at $\bx=\delta_3/2$
(Fig.~\ref{fig:G0}), with respect to the site-like case ($\bx=\bz$). Indeed,
while $\Im G^0 (\omega)$ is depressed with respect to the site-like case, as a
consequence of the finite extent of the gaussian pseudoatomic wavefunctions,
$\Re G^0 (\omega)$ attains a finite value at $\omega=\omega_\top$
(Fig.~\ref{fig:G0}). This implies the existence of bound states above the
conduction band only for $U_0 \gtrsim 8t$, while bound states below the valence
band still exists for any $U_0 <0$. On the other hand, resonances between the
two Van~Hove singularites close to $\omega=0$ are possible for $2<
U_0/t <20$, \emph{i.e.} within a finite range of positive values of the
impurity strength. 

An expansion of $\Re G^0$ in Eq.~(\ref{eq:rescond}), where now $\bx=\delta_3/2$,
leads to the estimate
\begin{equation}
\frac{t}{U_0} \approx \left( A_b + 2 s \right) \phi^2 (\delta_3/2) 
\label{eq:rescond2}
\end{equation}
for the impurity potential required to generate a bound state at
$\omega=0$, where $A_b \simeq 0.67$ (Appendix~\ref{app:tightbinding}). As in
the site-like case, one can recognize a term due to the asymmetry between the
two bands ($s\neq0$).

Fig.~\ref{fig:bondtosite} compares the LDOS for a site-like (top row panels) and
a bond-like (bottom row panels) impurity, both evaluated on the same position as
the impurity (left column panels) and on a nearest neighbor lattice site (right
column panels), for several values of the potential strength. One finds quite a
different behavior in the two cases. As $U_0$ increases towards $\tilde{U}_0$,
\emph{i.e.} the value of the impurity strength giving rise to a bound state at
$\omega=0$, the LDOS on the impurity site, and the weight of the Van~Hove
singularities, decreases, as expected. On the other hand, the bound state at
$\omega=0$ becomes sharper and more pronounced in the bond-like case, whereas it
becomes suppressed in the site-like case. Another remarkable difference is the
presence, in the bond-like case, of a wide resonant state in the conduction band
for $4t\lesssim U_0 \lesssim 8t$, which is completely absent in the site-like
case. This can be traced back to the different $\omega$-dependence of $\Re G^0$
in Eq.~(\ref{eq:rescond}) in the two cases. Fig.~\ref{fig:bondtosite} also
compares the LDOS for the site-like and bond-like cases, but now evaluated on a
nearest neighbor lattice site to the impurity position. Again, the Van~Hove
singularities become smoother, as $U_0$ increases towards $\tilde{U}_0$. The
suppression of the singularities due to an increase in the potential strength is
enhanced in the bond-like case, than in the site-like case.

\begin{figure}[t]
\begin{center}
\begin{minipage}[c]{0.48\columnwidth}
\begin{center}
\includegraphics[height=\textwidth,angle=-90]{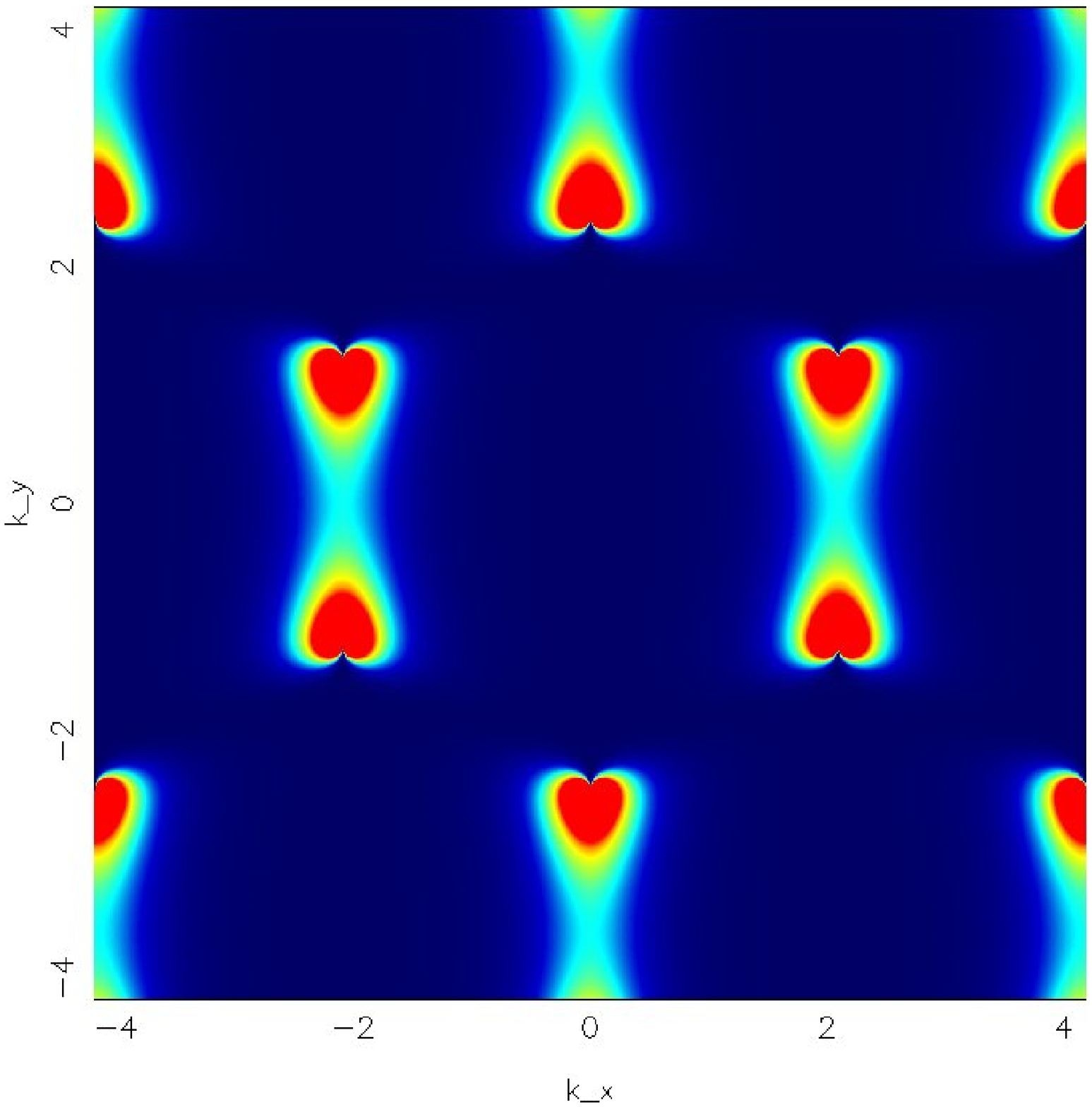}
\end{center}
\end{minipage}
\begin{minipage}[c]{0.48\columnwidth}
\begin{center}
\includegraphics[height=\textwidth,angle=-90]{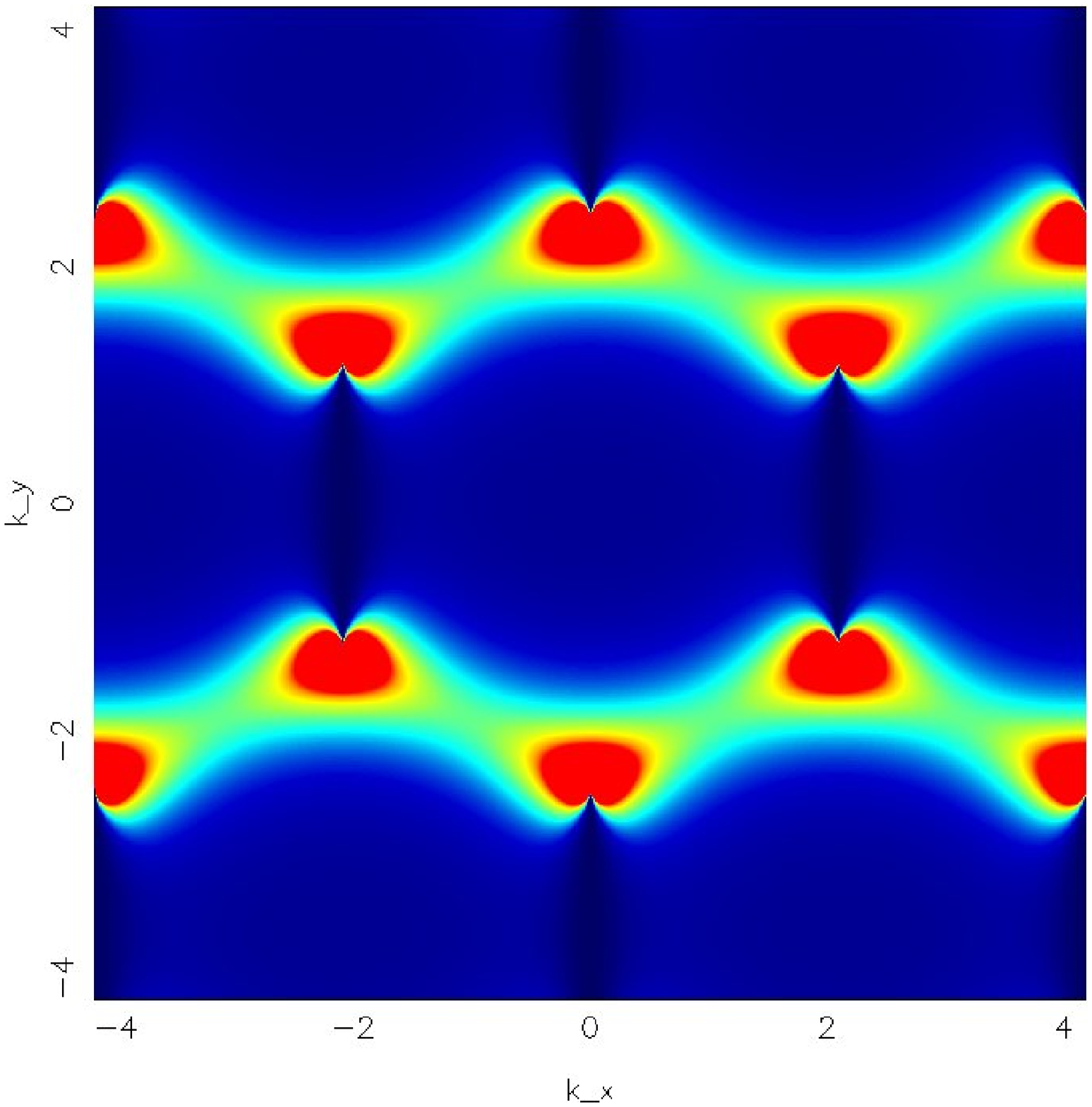}
\end{center}
\end{minipage}
\caption{(Color online) Contour plots of the LDOS in momentum space,
$\rho_\lambda (\bk,\omega)$,  Eq.~(\ref{eq:LDOSk}), for the valence
($\lambda=1$, left panel) and conduction band ($\lambda=2$, right panel). Here,
we are considering a bond-like impurity with $U_0 = 5.1t$, thus giving rise to a
bound state at $\omega=0$.}
\label{fig:rhow0bond}
\end{center}
\end{figure} 

Finally, Fig.~\ref{fig:rhow0bond} shows the LDOS in momentum space,
Eq.~(\ref{eq:LDOSk}), in the case of a bond-like impurity at $\bx=\delta_3/2$,
for the valence and conduction bands. Fig.~\ref{fig:rhow0bond} refers to a
potential strength of $U_0 = 5.1t$, thus giving rise to a bound state at
$\omega=0$. Similar pictures, but rotated of multiples of $2\pi/3$, would be
obtained in the other, inequivalent, bond-like positions. As in the site-like
case, Fig.~\ref{fig:rhow0onsite}, one finds that the points in $\bk$-space
providing the largest contribution to $\rho_\lambda (\bk,\omega)$ are those
closer to the Dirac points, but now with a reduced symmetry. In particular,
$\rho_\lambda (\bk,\omega)$ is not invariant with respect to transformations of
the $C_{6v}$ point group because of the squared modulus of the pseudoatomic
wavefunctions in Eq.~(\ref{eq:LDOSk}).

\subsection{Hollow-like impurities}
\label{ssec:hollow}

\begin{figure}[t]
\centering
\includegraphics[height=0.8\columnwidth,angle=-90]{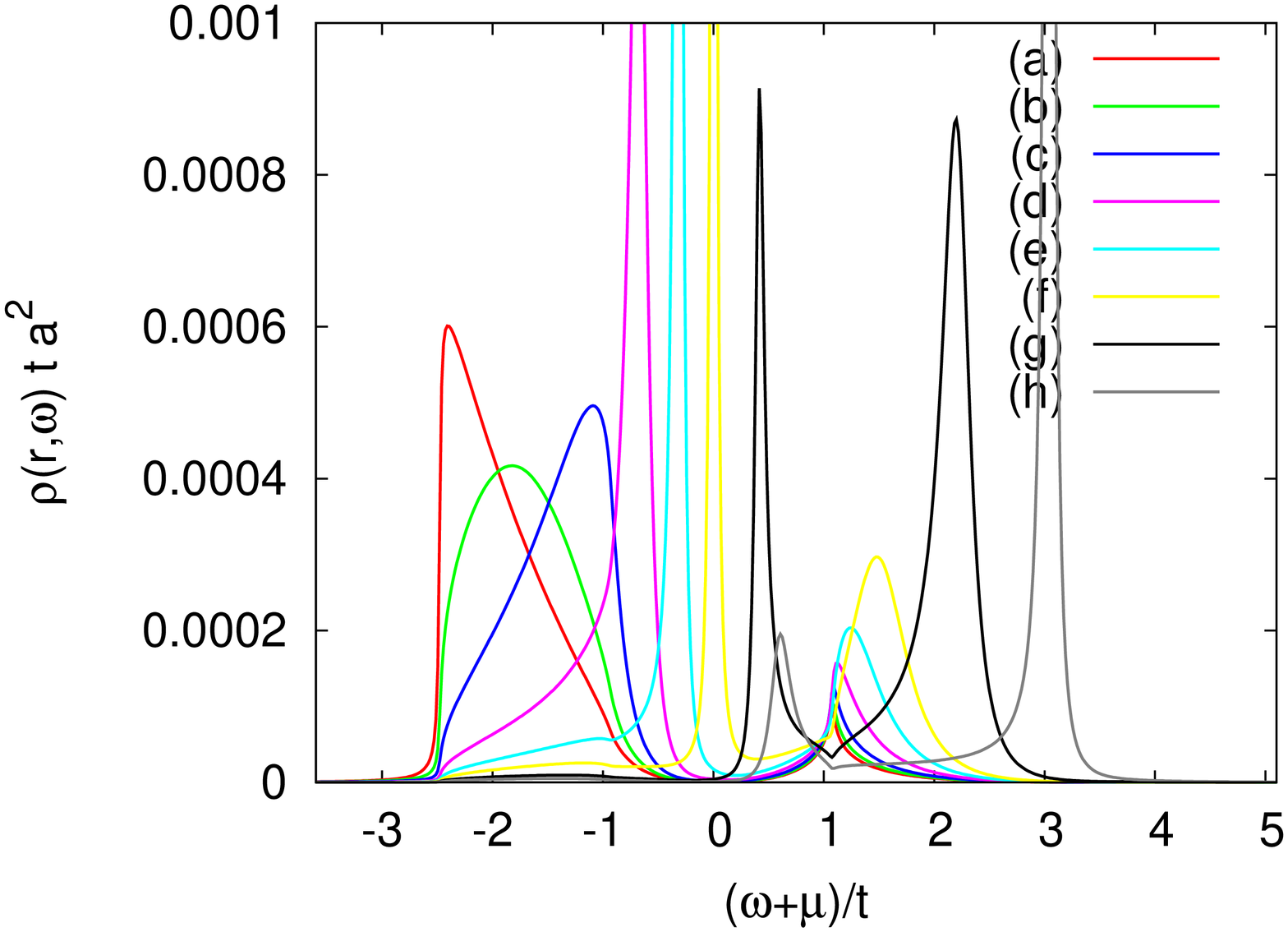}
\includegraphics[height=0.8\columnwidth,angle=-90]{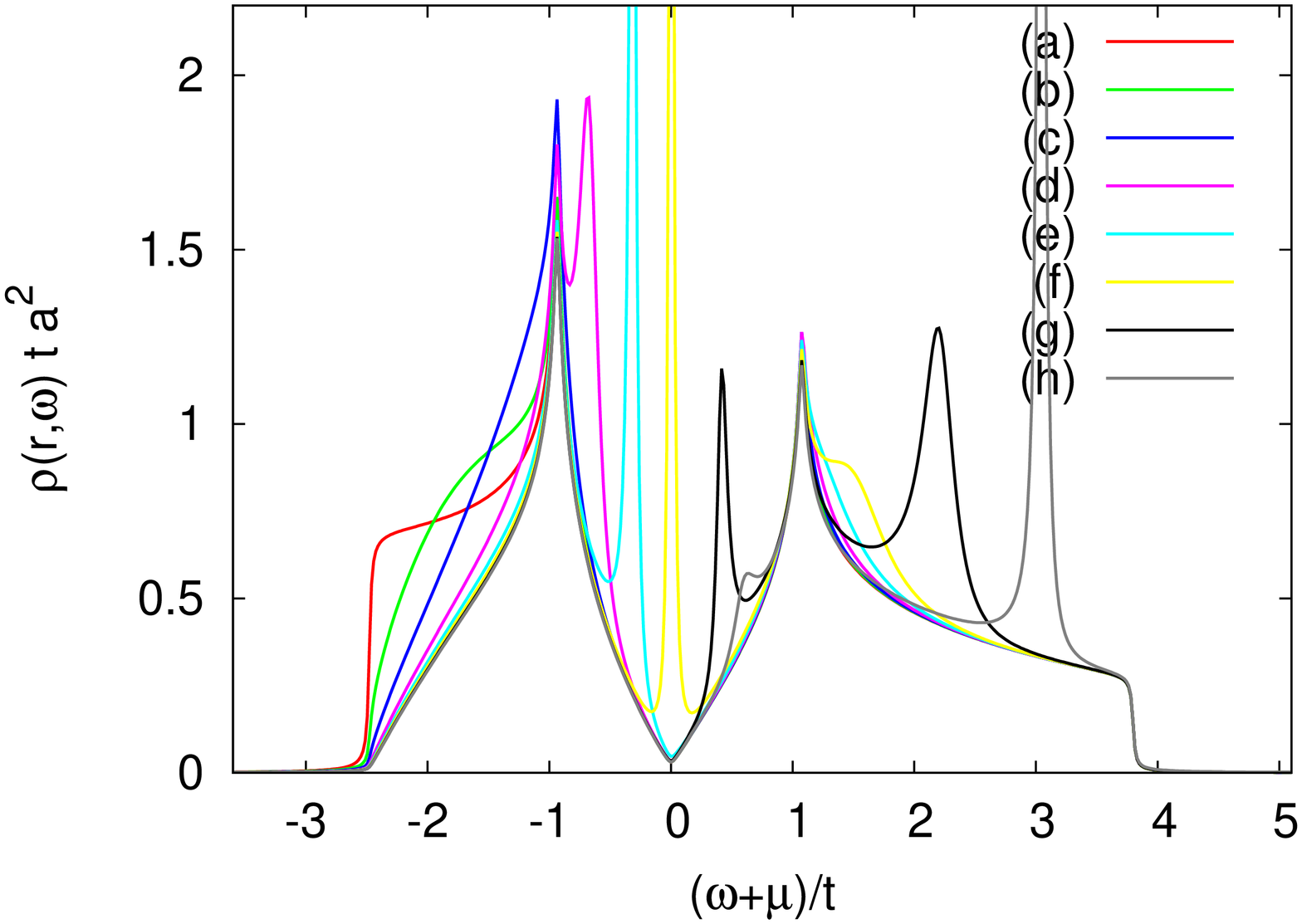}
\caption{(Color online) Local density of states for a hollow-like impurity
($\bx=-\delta_3$) on
the same site as the impurity (top
panel, $\br=\bx$), and on a nearest neighbor lattice site (bottom panel,
$\br=\bz$). 
The potential strengths are 
(a) $U_0 =0$; 
(b) $U_0 =0.10 \tilde{U}_0$; 
(c) $U_0 =0.25 \tilde{U}_0$;
(d) $U_0 =0.50 \tilde{U}_0$;
(e) $U_0 =0.75 \tilde{U}_0$;
(f) $U_0 = \tilde{U}_0$;
(g) $U_0 = 1.50 \tilde{U}_0$;
(h) $U_0 =2.00 \tilde{U}_0$,
where $\tilde{U}_0$ is the value of $U_0$ yielding a bound state at $\omega=0$.}
\label{fig:hollowtosite}
\end{figure}

The last case considered here corresponds to having a single impurity located at
the center of an hexagon plaquette, $\bx=-\delta_3$, say. This is the highest
symmetry position in the carbon honeycomb lattice, and indeed the point symmetry
$D_{6h}$ is preserved. Inspection of Fig.~\ref{fig:G0} for the
$\omega$-dependence of $\Re G^0 (\omega)$ and $\rho^0 (\omega)$ shows that the
unperturbed LDOS is severely depressed (some three orders of magnitude lower)
than the LDOS in the site-like case. Moreover, as a consequence of the overall
behavior of $\Re G^0 (\omega)$, one has a bound state below the valence band for
all negative values of $U_0$, whereas one has a bound state above the conduction
band for relatively large positive values of the impurity strength, $U_0 \gtrsim
10^4 t$. On the other hand, the relatively low value of $\rho^0 (\omega)$ allows
the formation of well-resolved resonant states close to $\omega=0$, for
$1.5\cdot 10^3 t \lesssim U_0 \lesssim 1.1 \cdot 10^4 t$. Expanding $\Re G^0$ in
Eq.~(\ref{eq:rescond}), where now $\bx=-\delta_3$, yields in this case the estimate
\begin{equation}
\frac{t}{U_0} \approx \left( A_h + 2 B_h s \right) \phi^2 (\delta_3)
\label{eq:rescond3}
\end{equation}
for the impurity potential required to generate a bound state at
$\omega=0$, where $A_h \simeq 2.35$ and $B_h =3$
(Appendix~\ref{app:tightbinding}). As in the previous cases, the main term persists
also in the limit of perfect band symmetry ($s=0$).

Fig.~\ref{fig:hollowtosite} shows the LDOS for a hollow-like impurity, both on
top of the impurity site, and on an adjacent lattice site, for several potential
strengths. At variance from the previous two cases, it is apparent that resonant
states are sharper in the nearest neighbor site, than on top of the impurity
position. Analogously to the bond-like case, there are resonant states
developing in the high conduction band, which are however better defined.

\begin{figure}[t]
\begin{center}
\begin{minipage}[c]{0.48\columnwidth}
\begin{center}
\includegraphics[height=\textwidth,angle=-90]{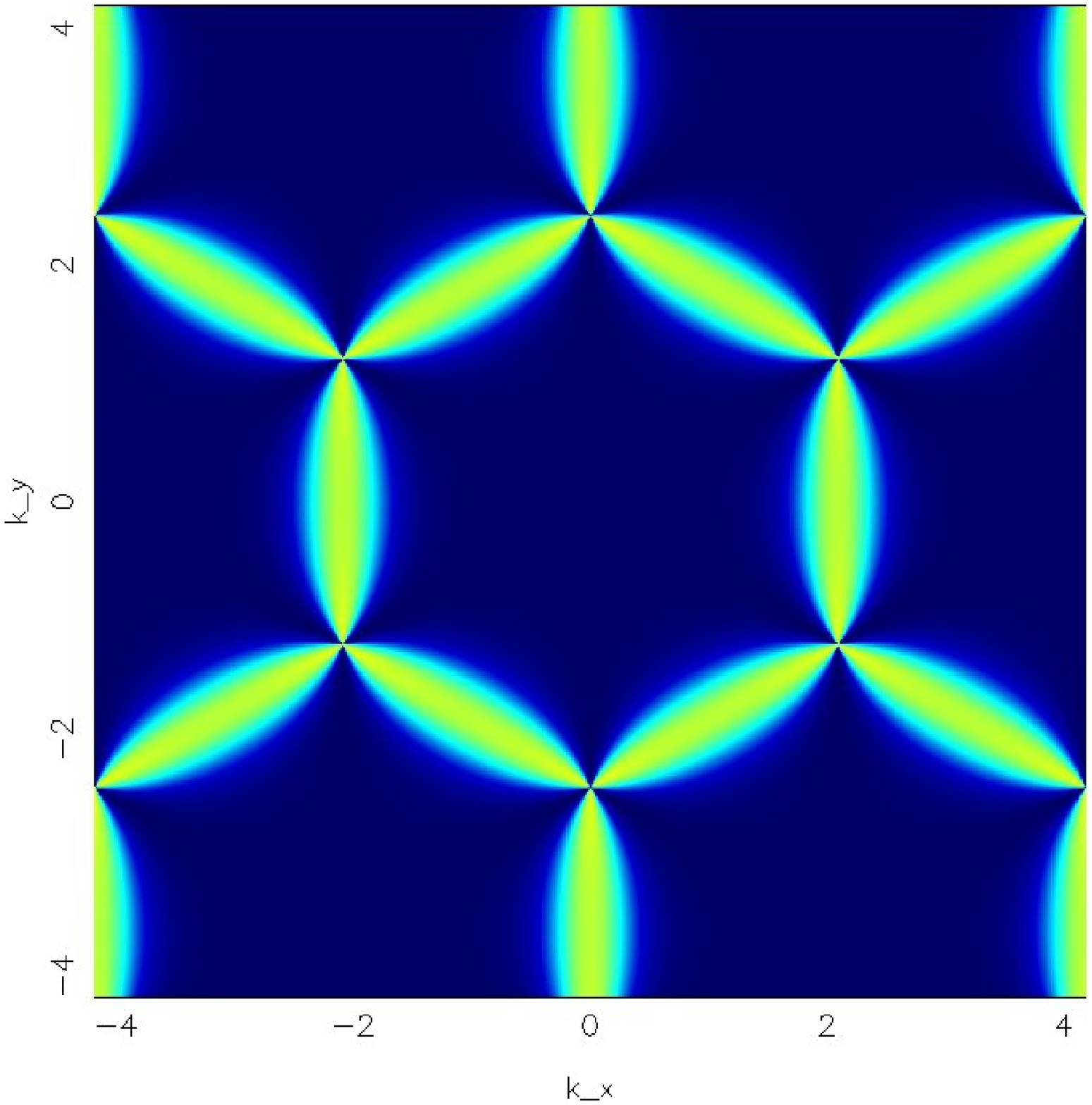}
\end{center}
\end{minipage}
\begin{minipage}[c]{0.48\columnwidth}
\begin{center}
\includegraphics[height=\textwidth,angle=-90]{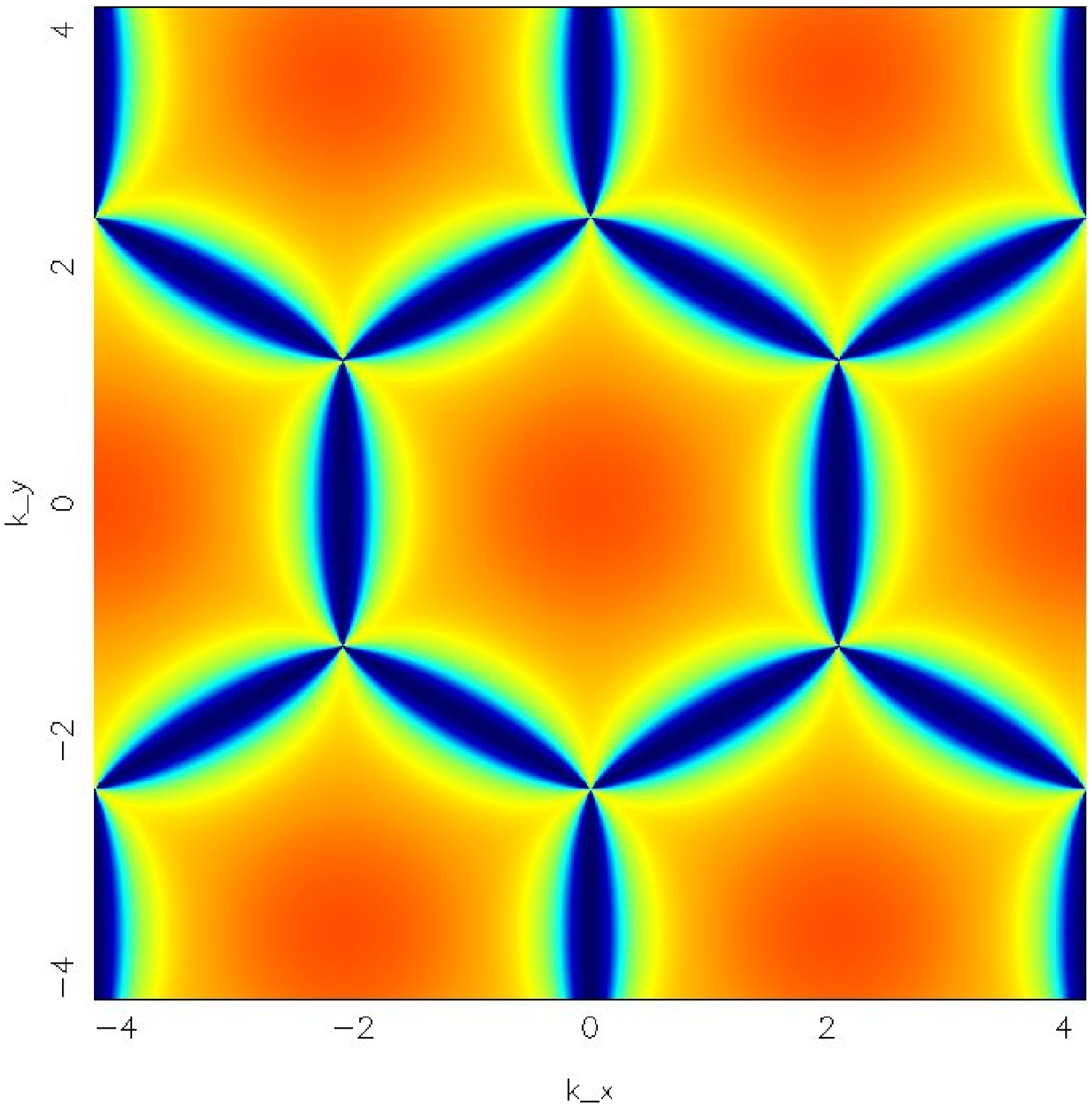}
\end{center}
\end{minipage}
\caption{(Color online) Contour plots of the LDOS in momentum space,
$\rho_\lambda (\bk,\omega)$,  Eq.~(\ref{eq:LDOSk}), for the valence
($\lambda=1$, left panel) and conduction band ($\lambda=2$, right panel). Here,
we are considering a hollow-like impurity with $U_0 = 3.8\cdot 10^4 t$, thus giving rise to a
bound state at $\omega=0$.}
\label{fig:rhow0hollow}
\end{center}
\end{figure} 

Fig.~\ref{fig:rhow0hollow} shows the LDOS in momentum space, for a hollow-like
impurity giving rise to a bound state at $\omega=0$, Eq.~(\ref{eq:rescond3}). At
variance with the previous cases, one may notice that the $\bk$-states
contributing most importantly to $\rho_\lambda (\bk,\omega)$ are the same as
those involved in building up the unperturbed LDOS. Indeed, in the conduction
band ($\omega>0$), the largest contributions to $\rho^0
(-\delta_3,-\delta_3,\omega)$ come from the Van~Hove singularities and the
centers of the sides of the first Brillouin zone. Similarly, in the valence band
($\omega<0$), the largest contributions to $\rho^0 (-\delta_3,-\delta_3,\omega)$
come from the band bottom, \emph{i.e.} from $\bk$-points close to the $\Gamma$
point. This can be ultimately be traced back to the extended width of the
gaussian pseudoatomic wavefunction here employed.

\section{Many impurities}
\label{sec:many}

While single impurity effects are in principle observable through STM
measurements \cite{Ishigami:07,Stolyarova:07,Geringer:09}, real samples usually
contain a sizeable amount of impurities, which are responsible of sensible
modifications of both thermodynamic and transport properties. Therefore, we will
here exploit the results of Sec.~\ref{sec:single} for a single impurity, to
study the effect of $N_{imp}$ impurities on a graphene monolayer. We will assume
that (i) the position of all impurities differ by a vector of the direct
lattice; in other words, there is a preferential kind of impurity location,
\emph{i.e.}  all impurities  are either site-like, bond-like, or hollow-like,
according to the classification given in the Sec.~\ref{sec:single}; (ii)
impurities are independent, \emph{i.e.} the average distance between two
impurities is larger than the quasiparticle coherence length, so that
interference effects can be neglected; (iii)  their number is sufficiently large
($N_{imp}\gg1$), so that their effect is appreciable on bulk properties in the
thermodynamic limit, but the impurities are sufficiently diluted
($n_{imp}=N_{imp}/N\ll1$). In these limits, while a standard averaging procedure
over the position configurations of the impurities restores the translational
invariance of the Green's function,
\begin{equation}
\mathcal{G}^{imp}_{\lambda\lambda^\prime} (\bk,\bk^\prime,i\omega_n) =
\delta_{\bk\bk^\prime} \mathcal{G}^{imp}_{\lambda\lambda^\prime}
(\bk,i\omega_n),
\end{equation}
the eigenstates of the pure Hamiltonian are expected to acquire a
finite lifetime induced by disorder. This can be formally achieved by relating
$\mathcal{G}^{imp} (\bk,i\omega_n)$, now a matrix with respect to the band
indices, to the single-impurity Green's function $\mathcal{G}^0 (\bk,i\omega_n)$
discussed in Sec.~\ref{sec:single} through a Dyson's equation analogous to
Eq.~(\ref{eq:Dyson}), but now involving the proper self energy matrix \cite{Bruus:04} $\Sigma
(\bk,i\omega_n)$ 
\begin{eqnarray}
\mathcal{G}^{imp} (\bk,i\omega_n) &=& \mathcal{G}^0 (\bk,i\omega_n) \nonumber\\
&&\hspace{-1truecm}+ 
\mathcal{G}^0 (\bk,i\omega_n) \Sigma (\bk,i\omega_n)
\mathcal{G}^{imp} (\bk,i\omega_n) .
\label{eq:DysonGimp}
\end{eqnarray}
Within the full Born approximation (FBA) \cite{Bruus:04}, which is valid in the
limit of small impurity concentration, $n_{imp}^2 \ll n_{imp}$, one finds
\begin{equation}
\Sigma_{\lambda\lambda^\prime} (\bk,i\omega_n) =
n_{imp} \frac{V_0 \check{\psi}^\ast_{\bk\lambda} (\bx)
\check{\psi}_{\bk\lambda^\prime} (\bx)}{1-\frac{V_0}{N}
\sum_{\bq\lambda^{\prime\prime}} |\check{\psi}_{\bk\lambda^{\prime\prime}}
(\bx)|^2 \mathcal{G}^0_{\lambda^{\prime\prime}} (\bq,i\omega_n )} ,
\label{eq:Gimp}
\end{equation}
where we are assuming that all impurities occupy equivalent lattice positions.
Comparing Eq.~(\ref{eq:Gimp}) with Eqs.~(\ref{eq:Tmatrix}) and (\ref{eq:G0}), it
is possible to relate the proper self-energy within the FBA to the $T$-matrix
for the same kind of impurity, through
\begin{equation}
\Sigma_{\lambda\lambda^\prime} = N_{imp} T_{\lambda\lambda^\prime} (\bx;
\bk,\bk^\prime,i\omega_n).
\label{eq:SigmaT}
\end{equation}
Eq.~(\ref{eq:SigmaT}) therefore enables us to generalize most of the results
derived in Sec.~\ref{sec:single} to the case of many impurities, all located
within a preferential class of lattice positions. 


\subsection{LDOS}
\label{ssec:manyldos}

We begin by discussing the effect of many impurities on the local density of
states in reciprocal space. This can be obtained by inverting Eq.~(\ref{eq:DysonGimp}) for
$\mathcal{G}^{imp} (\bk,i\omega_n )$ and then performing the usual analytical
continuation. Most properties can be derived by describing the behavior of the
analytically-continued proper self-energy, which in all of the three cases of
interest can be written as
\begin{equation}
\Sigma(\bk,\omega) = n_{imp} \frac{V_0 W(\bk)}{1-V_0 G^0 (\bx,\bx,\omega)} ,
\end{equation}
where $\bx=\bz$, $\delta_3/2$, or $-\delta_3$ in the site-like, bond-like, or
hollow-like case, respectively, and $W(\bk)$ is a matrix form factor explicitly
given by
\begin{subequations}
\begin{eqnarray}
\label{eq:Ws}
W^{(s)}_{\lambda\lambda^\prime} (\bk) &=&
\frac{1}{2} [
\check{\psi}^\ast_{\bk\lambda} (\bz)
\check{\psi}_{\bk\lambda^\prime} (\bz)
+
\check{\psi}^\ast_{\bk\lambda} (\delta_3)
\check{\psi}_{\bk\lambda^\prime} (\delta_3)
] \\
\label{eq:Wb}
W^{(b)}_{\lambda\lambda^\prime} (\bk) &=& \frac{1}{3} \sum_{\ell=1}^3
\check{\psi}^\ast_{\bk\lambda} (\delta_\ell/2)
\check{\psi}_{\bk\lambda^\prime} (\delta_\ell/2) \\
\label{eq:Wh}
W^{(h)}_{\lambda\lambda^\prime} (\bk) &=& 
\check{\psi}^\ast_{\bk\lambda} (-\delta_3)
\check{\psi}_{\bk\lambda^\prime} (-\delta_3)
\end{eqnarray}
\end{subequations}
in the site-like, bond-like, and hollow-like cases, respectively. We are here
assuming that the impurities are equally distributed among the $A$ and $B$
sites, in the site-like case, and among the three classes of $\sigma$ bonds, in
the bond-like case.

Both in the site-like and in the bond-like cases, direct inspection of the
solution of Eq.~(\ref{eq:DysonGimp}) shows that $G^{imp} (\bk,\omega)$ is nearly
diagonal in the diluted limit ($n^2_{imp}\ll n_{imp}$). Therefore, an eigenstate
of the unperturbed Hamiltonian labelled by wavevector $\bk$ and band index
$\lambda$ acquires a finite lifetime $\tau_{\bk\lambda}$, which \emph{e.g.} in
the site-like case and in the limit of low LDOS is given by
\begin{equation}
\tau_{\bk\lambda}^{-1} \approx \pi n_{imp} V_0^2 W_{\lambda\lambda}  (\bk)
\rho(\bz, \omega=\xi_{\bk\lambda} ),
\end{equation}
where $\rho(\bx,\omega)$ is the LDOS with a single impurity,
Eq.~(\ref{eq:LDOSx}).

For impurity potentials close to the condition for a well-defined resonance at
$\omega\approx0$ in the single-impurity case, Eq.~(\ref{eq:rescond}), in the dilute
limit, one gets for the LDOS in reciprocal space close to a Dirac point
\begin{equation}
\rho_\lambda (\bk,\omega) \approx \frac{A_{\bk\lambda}}{[\omega-\xi_{\bk\lambda} + B_{\bk\lambda}
(\omega-\xi_{\bk\bar{\lambda}})]^2 + \pi^2 A_{\bk\lambda}^2} ,
\end{equation}
where $\bar{\lambda}=2$ when $\lambda=1$, and $\bar{\lambda}=1$ when
$\lambda=2$, and
\begin{subequations}
\begin{eqnarray}
A_{\bk\lambda} &=& 
n_{imp} V_0^2 W_{\lambda\lambda} (\bk)
\rho_\lambda (\bz,\omega) \nonumber\\
&&\times \left[
1 - \frac{W_{12}(\bk)W_{21}(\bk)}{W_{11}(\bk)W_{22}(\bk)} \right],\\
B_{\bk\lambda} &=&
\frac{W_{12}(\bk)W_{21}(\bk)}{W_{\bar{\lambda}\bar{\lambda}}^2 (\bk)} .
\end{eqnarray}
\end{subequations}
The behavior of the LDOS in reciprocal space is therefore quite different from
the unperturbed case, which would be characterized by a Dirac delta peaked along
closed contours around the Dirac points.

The case in which all impurities are located in a hollow-like position,
Eq.~(\ref{eq:Wh}), is quite different from the previous two cases,
Eqs.~(\ref{eq:Ws}) and (\ref{eq:Wb}). This is due to the fact that the form
factors $W(\bk)$ defined in Eq.~(\ref{eq:Wh}) fulfill the additional identity
$W_{11} (\bk) W_{22} (\bk) - W_{12}(\bk)W_{21}(\bk)=0$. 
\modified{%

While the Born approximation holds for low impurity concentration, $n_{imp}^2
\ll n_{imp}$, one has to distinguish two regimes. For moderately large impurity
concentrations, $n_{imp} \gg t\, |\Im G^0 (-\delta_3 ,\omega_{res} )|$, for an
impurity potential close to what would be a resonance in the single-impurity
case, one finds a nearly diagonal Green's function, whose nonzero matrix
elements are given by
\begin{eqnarray}
G^{imp}_{\lambda\lambda} &\approx& 
\left[
\omega - \xi_{\lambda\bk} +
\frac{W_{\lambda\lambda}(\bk)}{W_{\bar{\lambda}\bar{\lambda}} (\bk)}
(\omega - \xi_{\bar{\lambda}\bk}) \right.\nonumber\\
&&\left. + \frac{i}{\pi}
\frac{W_{\lambda\lambda}(\bk)}{W_{\bar{\lambda}\bar{\lambda}}^2(\bk)}
\frac{(\omega-\xi_{\bar{\lambda}\bk})^2}{n_{imp} V_0^2 \rho(-\delta_3,\omega)}
\right]^{-1} .
\label{eq:Gimpapprox}
\end{eqnarray}
In the same limit, for an impurity potential giving rise to a resonance at
exactly $\omega=0$, the situation is even more dramatic, since
\begin{equation}
G^{imp}_{\lambda\lambda} \approx \left[ -\left(\xi_{\lambda\bk} +
\frac{W_{\lambda\lambda}(\bk)}{W_{\bar{\lambda}\bar{\lambda}}
(\bk)}\xi_{\bar{\lambda}\bk} \right)+ i\eta \right]^{-1} ,
\end{equation}
where $\eta$ is a positive infinitesimal.
On the other hand, for $n_{imp} \ll t\, |\Im G^0 (-\delta_3 ,\omega_{res} )|$,
one finds
\begin{equation}
G^{imp}_{\lambda\lambda} \approx
\left[
\omega - \xi_{\lambda\bk} +
i\pi n_{imp} V_0^2 \rho(-\delta_3 ,\omega) W_{\lambda\lambda}(\bk) \right]^{-1}.
\label{eq:Gimpapprox2}
\end{equation}
The different behavior with respect to the previous two cases follows from the
fact that in the hollow-like case the Born approximation performs an average
with respect to impurity positions all of the same kind, at variance,
\emph{e.g.} with the site-like case, where impurities can be added either in the
$A$ or in the $B$ sublattices. This is likely to produce an additional dephasing
among contributions arising from inequivalent lattice positions in the site-like
or bond-like cases, with respect to the hollow-like case, thereby resulting in
an increasing inverse lifetime with increasing impurity concentration, although
only in the moderately low impurity concentration, Eq.~(\ref{eq:Gimpapprox}).
One however recovers the physically expected behavior at low impurity
concentrations, \emph{i.e.} a vanishing inverse lifetime with increasing
impurity concentration, Eq.~(\ref{eq:Gimpapprox2}).}

\subsection{Conductivity}
\label{ssec:conductivity}

We end this section by considering the effect of many, short-range impurities on
the conductivity. As described above, we are mainly concerned with the case in
which all impurities are located in the same class of lattice positions, and we
will here focus on the site-like case, even though, as described in
Sec.~\ref{ssec:manyldos}, the analysis presented here is actually more general,
as it qualitatively applies also to the bond-like case, at least for $\mu$
between the two Van~Hove singularities. Again, we will assume the dilute regime,
$n_{imp}^2 \ll n_{imp}$, so that the full Born approximation holds.

Within linear response theory, the conductivity $\sigma$ is related to the
current-current correlation function through a Kubo formula
\begin{equation}
\sigma_{lm} (\mu,T;\omega) = \frac{ie^2 n}{m\omega}\delta_{lm} +
\frac{i}{\hbar\omega N A_{cell}} \tilde{\Pi}^\Ret_{lm} (0,0,\omega),
\end{equation}
where $n$ is the electron density, $\omega$ is the frequency of the external
electric field, $A_{cell}$ is the area of a primitive cell, and
$\tilde{\Pi}^\Ret_{lm} (\bk,\bq,\omega)$ is the $(l,m)$ component of the Fourier
transform of the retarded current-current correlation tensor. We are mainly
interested in the dissipative part of the conductivity tensor, \emph{i.e.} its
real part. For the longitudinal part $\sigma=\Re \sigma_{xx}$, one has
\begin{equation}
\sigma (\mu,T;\omega) = - \frac{1}{\hbar\omega N A_{cell}} \Im
\tilde{\Pi}^\Ret_{xx} (0,0,\omega),
\label{eq:sigma}
\end{equation}
where $\tilde{\Pi}^\Ret_{lm}$ is the retarded version of
\begin{equation}
\tilde{\Pi}_{lm} (\bk,\bq,\tau) = - \langle T_\tau [ \tilde{J}^\nabla_l
(\bk,\tau) \tilde{J}^\nabla_m (\bq,0) ] \rangle ,
\end{equation}
and $\tilde{J}^\nabla_l (\bk,\tau)$ denote the Fourier transform of the
paramagnetic component of the current density vector, at the imaginary time
$\tau$. Expanding $\tilde{J}^\nabla_l (\bk,\tau)$ in terms of the eigenstates of
the unperturbed Hamiltonian, and making use of the results of
Appendix~\ref{app:current}, one finds
\begin{widetext}
\begin{equation}
\tilde{\Pi}_{xx} (0,0,\tau) =
e^2 \frac{t^2 a^2}{\hbar^2}
\sum_{{\bk\bk^\prime}\atop{\lambda\lambda^\prime\eta\eta^\prime}}
h_{x,\lambda\eta}(\bk) h_{x,\lambda^\prime\eta^\prime}(\bk^\prime) 
\langle T_\tau [ c^\dag_{\bk\lambda}(\tau^+) c_{\bk\eta}(\tau)
c^\dag_{\bk^\prime\lambda^\prime}(0^+)c_{\bk^\prime\eta^\prime}(0)]\rangle,
\end{equation}
where the matrix elements $\bh_{m,\lambda\eta}(\bx)$ are defined in
Appendix~\ref{app:current}. One is now in the position to make use of Wick's
theorem. We further make the approximation to treat the one-body Green's
functions within the FBA, and perform the required analytical continuation, to
obtain
\begin{eqnarray}
\tilde{\Pi}^\Ret_{xx} (0,0,\omega) &=& 2e^2 \frac{t^2 a^2}{\hbar^2}
\sum_{{\bk}\atop{\lambda\lambda^\prime\eta\eta^\prime}}
h_{x,\lambda\eta}(\bk) h_{x,\lambda^\prime\eta^\prime}(\bk)
\int_{-\infty}^\infty \frac{d\zeta}{2\pi i\hbar}
\left\{
[\nF (\zeta+\omega) - \nF(\zeta)]
G^{imp\ast}_{\lambda\eta^\prime}(\bk,\zeta)
G^{imp}_{\eta\lambda^\prime}(\bk,\zeta+\omega) \right.\nonumber\\
&&\left.+\nF(\zeta)
G^{imp}_{\eta^\prime\lambda}(\bk,\zeta)G^{imp}_{\eta\lambda^\prime}(\bk,\zeta+\omega)
-\nF(\zeta+\omega)G^{imp\ast}_{\lambda\eta^\prime}(\bk,\zeta)
G^{imp\ast}_{\lambda^\prime\eta}(\bk,\zeta+\omega) \right\},
\end{eqnarray}
\end{widetext}
where $\nF(\omega)$ is the Fermi function, and the factor of 2 takes into
account for spin degeneracy.

We next make a further approximation, \emph{i.e.} we assume that the impurity
Green's functions are diagonal in the band index,
$G^{imp}_{\lambda\lambda^\prime}
(\bk,\omega) \approx \delta_{\lambda\lambda^\prime} G^{imp}_{\lambda\lambda}
(\bk,\omega)$. Such an approximation is justified in the dilute limit, and
amounts to treat the effect of disorder as a perturbation to the pure case,
whose main effect is that of adding a finite lifetime to the eigenstates of the
unperturbed Hamiltonian.

\begin{figure}[b]
\centering
\includegraphics[height=0.8\columnwidth,angle=-90]{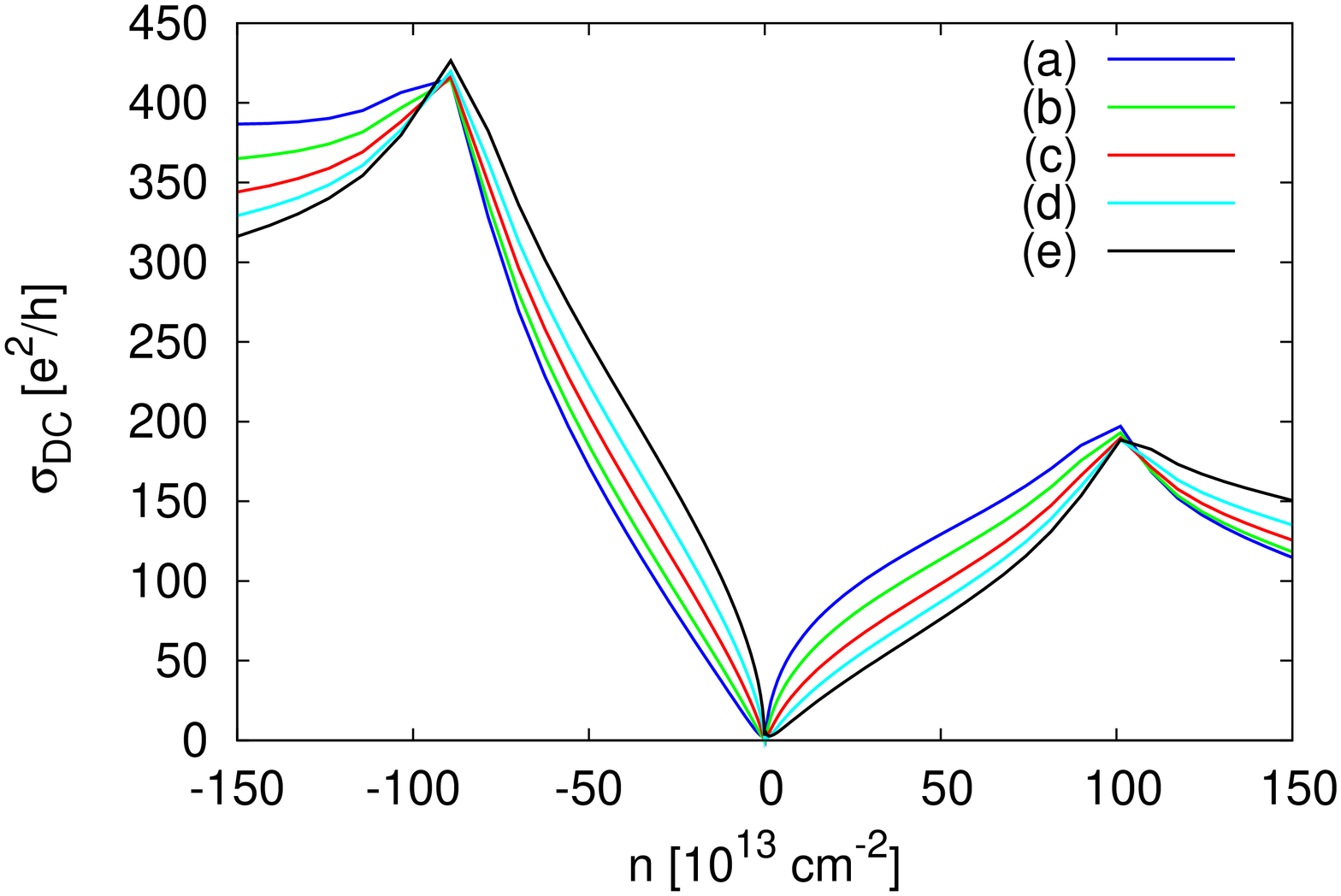}
\includegraphics[height=0.8\columnwidth,angle=-90]{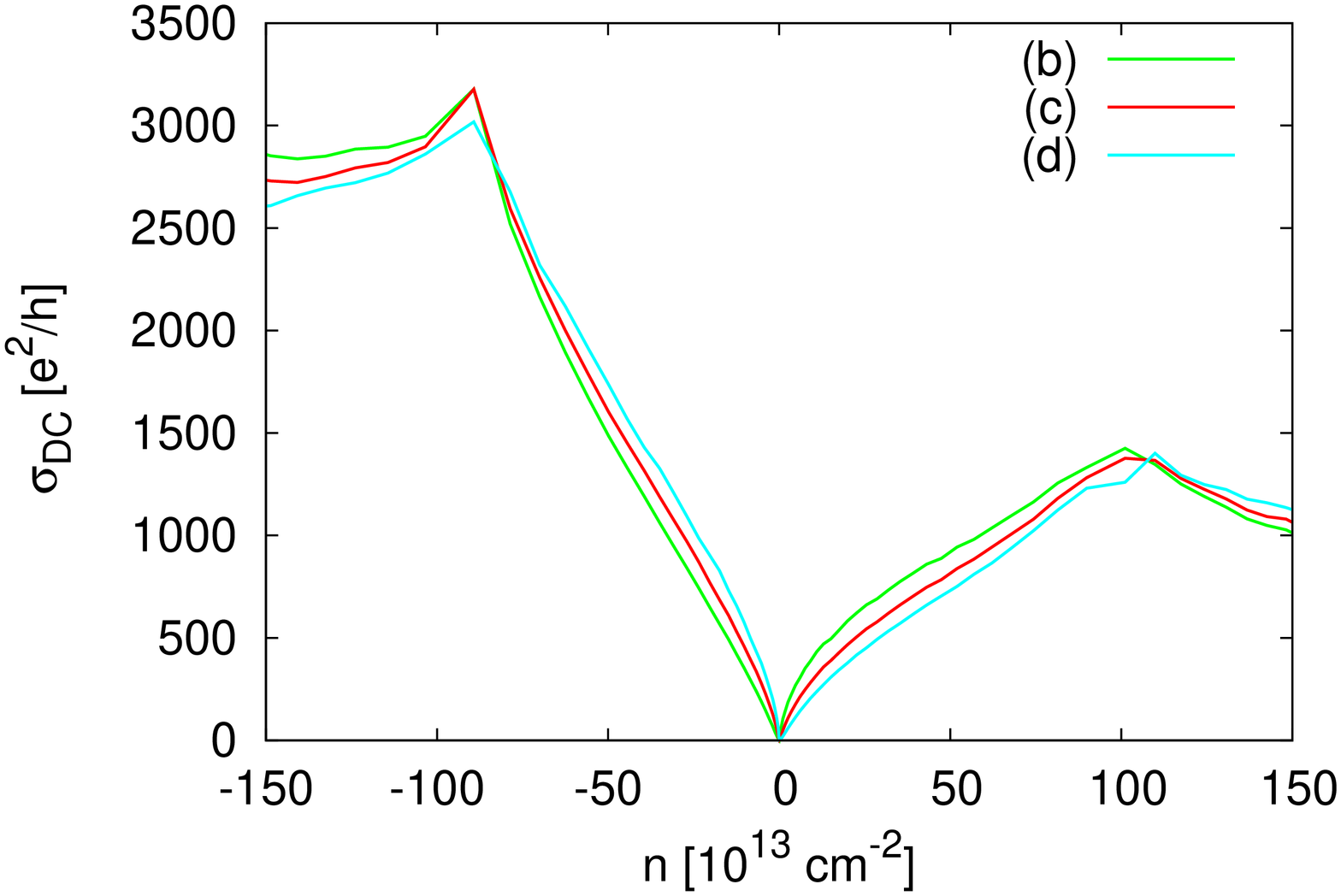}
\caption{(Color online) Conductivity $\sigma_{\mathrm{DC}}$ as a function of the carrier
concentration $n$ in graphene, in the presence of many site-like impurities.
The upper panel refers to a value $n_{imp}=10^{-2}$ of the impurity
concentration, while the lower panel is characterized by $n_{imp}=10^{-3}$. The
impurity potentials under consideration are 
(a) $U_0 = 0.35\tilde{U}_0$, 
(b) $U_0 = 0.50\tilde{U}_0$, 
(c) $U_0 = \tilde{U}_0$, 
(d) $U_0 = 10.00\tilde{U}_0$, 
(e) $U_0 = -\tilde{U}_0$, where $\tilde{U}_0$ is the potential strength giving
rise to a bound state at $\omega=0$.}
\label{fig:sigma}
\end{figure}

In the static limit ($\omega\to0$) and at $T=0$, Eq.~(\ref{eq:sigma}) yields the
conductivity as a function of the chemical potential, which can be decomposed in
an inter and intraband contribution, $\sigma_{\mathrm{DC}}(\mu) = \sigma_{inter}
(\mu) + \sigma_{intra}(\mu)$, given by
\begin{subequations}
\begin{eqnarray}
\label{eq:sigmainter}
\frac{\sigma_{inter} (\mu)}{\sigma_0} &=& 
-\frac{1}{\tau_0^2} \frac{1}{N}\sum_\bk
h_{x,12}(\bk) h_{x,21}(\bk) \nonumber\\
&&\times \Im G^{imp}_{11}(\bk,0) \Im G^{imp}_{22}(\bk,0)
\\
\label{eq:sigmaintra}
\frac{\sigma_{intra} (\mu)}{\sigma_0} &=& 
-\frac{1}{\tau_0^2}\frac{1}{N}\sum_{\bk,\alpha=1,2}
\left[h_{x,\alpha\alpha}(\bk) \Im G^{imp}_{\alpha\alpha}(\bk,0) \right]^2
\nonumber\\
\end{eqnarray}
\end{subequations}
where $\sigma_0 = \pi e^2/(2h)$ is proportional to the quantum of conductivity,
and $\tau_0^{-2} = 16t^2/(3\sqrt{3}\pi\hbar^2)$. One may expect that the
interband contribution, Eq.~(\ref{eq:sigmainter}), only becomes comparable with
the intraband contribution, Eq.~(\ref{eq:sigmaintra}), when $\mu\approx0$,
\emph{i.e.} when the valence and conduction bands overlap, owing to the
disorder-induced energy spread. Away from neutrality ($\mu=0$), and for a given
impurity potential strength $U_0$, an increase of the Fermi surface width
produces an increase of the conductivity. Such an increase is however rather
slow, close to the energy values where the LDOS with a single impurity is
maximum, where the backscattering due to the impurities is more effective. Such
a sublinear increase of $\sigma_{\mathrm{DC}}$ as a function of the carrier
concentration $n$ occurs for values of $U_0$ giving rise to resonant states
close to $\mu=0$ and does not depend on the value of $n_{imp}$. Such a behavior
is numerically confirmed in Fig.~\ref{fig:sigma}, for various values of
$n_{imp}$ and $U_0$, and is in good qualitative agreement with the experimental
results \cite{Bolotin:08}. The asymmetry between the particle ($n>0$) and hole
($n<0$) regimes is partly due to the band asymmetry ($s\neq0$), but is mainly
due to the effect of impurities, which is different depending on the sign of
$\mu$. Both in the valence and conduction bands, however, one observes the
occurrence of a maximum and then a decrease of $\sigma_{\mathrm{DC}}$ when $\mu$
attains the value corresponding to a Van~Hove singularity, where the Fermi
surface is maximally extended and traverses an electronic topological
transition. A comparison between the two panels in Fig.~\ref{fig:sigma} shows
that the nonmonotonic behavior of $\sigma_{\mathrm{DC}}$ is generic for all
impurity concentrations, but rather depends on the impurity potential $U_0$. The
similarity between Fig.~\ref{fig:sigma} and the concentration dependence of the
conductivity measured in suspended graphene after annealing surmises that
scattering due to short range impurities is relevant to determine the transport
properties of these graphene samples \cite{Tan:07,Stauber:08,Trushin:08}.

\section{Conclusions}
\label{sec:conclusions}

We have analyzed the effects of a single, localized impurity on the local
electronic properties of a graphene monolayer. Specifically, we have considered
an isolated impurity located on high-symmetry positions of the honeycomb
lattice, such as the site-like, bond-like, and hollow-like positions. While the
electronic properties of the pure system have been treated within the
tight-binding approximation, but allowing for asymmetry between valence and
conduction bands, the effect of the impurity has been modelled through a
gaussian pseudoatomic wavefunction, even though more general functional forms
have been taken into account.
\modified{%
Moreover, the tight binding scheme employed in this work does not suffer
from the `cone approximation' \cite{Peres:06,Skrypnyk:07}, thereby enabling one to treat both low and high
energies within the band with the same degree of accuracy.}

We have evaluated the local density of states as a function of energy on the
impurity site and on its nearest neighbor locations, and as a function of
wavevector in reciprocal space, close to a resonance. The latter may be of
relevance to interpret FTSTS measurements around an impurity. In particular, it
has been shown that the main contributions to the impurity-induced modification
of the LDOS come from wavevectors close to the Dirac point, in the site-like and
bond-like cases, while the same states are involved in determining the LDOS both
in the unperturbed case and in the case of a hollow-like impurity. 
\modified{%
Moreover, it has been suggested that FTSTS spectra can be used to distinguish
between different types of impurities, in particular as far as the impurity
potential range is concerned \cite{Bena:07}.}

We have determined semi-analytically the condition on the impurity strength for
having a bound state at $\omega=0$ in the three cases of interest. In
particular, in the site-like case, it is shown that the weight associated with a
bound state between the two Van~Hove singularities is larger for the LDOS on a
neighboring site, than on the impurity site itself. Such a behavior is analogous
to the one predicted for the $d$-density-wave state of high-$T_c$
superconductors, and can be attributed to the different contributions coming
from intra- and inter-valley impurity scattering.

Our results for the single-impurity case have been exploited to discuss the
effect of distributed impurities, all located in a preferential class of lattice
sites. Such a generalization has been derived within the full Born
approximation, and applies to the dilute limit. In particular, we have estimated
the quasiparticle lifetime associated to a finite impurity concentration, and
the behavior of the LDOS in reciprocal space. Within linear response theory, we
have also evaluated the static conductivity. One can again distinguish an intra-
and an inter-band contribution, the latter being sizeable only close to zero
carrier concentration, \emph{i.e.} when the two bands appreciably overlap.
Moving away from neutrality, one recovers a nonmonotonic dependence on the
carrier concentration, characterized by a sublinear increase close to $\mu=0$,
as is observed experimentally in suspended graphene samples after annealing.
Such a feature is generic, in the sense that it applies to all impurity
concentrations under study, and rather depends on the impurity strength.

\acknowledgments

The authors are indebted with Professor N.~H.~March for valuable discussions
over the general area embraced by the present work.

\appendix

\section{Dyson equation for separable impurity potential}
\label{app:Tmatrix}

\modified{%
Here, we briefly derive Dyson's equation, Eq.~(\ref{eq:DysonT}), in the case of 
a separable impurity potential, Eq.~(\ref{eq:Vsep}). Inserting our Ansatz for
the potential, Eq.~(\ref{eq:Vsep}), into Eq.~(\ref{eq:Dyson}), and iterating,
one may express the Green's function $\mathcal{G}_{\lambda\lambda^\prime}
(\bk,\bk^\prime,i\omega_n)$ as a series \cite{Bruus:04},
\begin{equation}
\mathcal{G}_{\lambda\lambda^\prime} (\bk,\bk^\prime,i\omega_n) =
\sum_{\ell=0}^\infty \mathcal{G}^{(\ell)}_{\lambda\lambda^\prime}
(\bk,\bk^\prime,i\omega_n) ,
\end{equation}
whose first and successive terms are given iteratively by
\begin{widetext}
\begin{subequations}
\label{eq:Giter}
\begin{eqnarray}
\mathcal{G}_{\lambda\lambda^\prime}^{(0)}
(\bk,\bk^\prime,i\omega_n) &=& \delta_{\lambda\lambda^\prime}
\delta_{\bk\bk^\prime} \mathcal{G}^{(0)}_\lambda (\bk,i\omega_n) \\
\mathcal{G}_{\lambda\lambda^\prime}^{(\ell)}
(\bk,\bk^\prime,i\omega_n) &=&
\mathcal{G}_{\lambda\lambda^\prime}^{(\ell-1)}
(\bk,\bk^\prime,i\omega_n) 
+ V_0 \sum_{\bq\lambda^{\prime\prime}}
\mathcal{G}_{\lambda\lambda^{\prime\prime}}^{(\ell-1)} 
(\bk,\bq,i\omega_n) 
\psi^\ast_{\bq\lambda^{\prime\prime}} (\bx) \psi_{\bk^\prime\lambda^\prime}
(\bx) \mathcal{G}^{(0)}_{\lambda^\prime} (\bk^\prime ,i\omega_n) .
\end{eqnarray}
\end{subequations}
This leads to the series
\begin{eqnarray}
\mathcal{G}_{\lambda\lambda^\prime} (\bk,\bk^\prime,i\omega_n) &=&
\delta_{\lambda\lambda^\prime} \delta_{\bk\bk^\prime} \mathcal{G}^{(0)}_\lambda (\bk,i\omega_n)
\nonumber\\
&&+ V_0 \mathcal{G}^{(0)}_\lambda (\bk,i\omega_n) 
\psi^\ast_{\bk\lambda} (\bx) \psi_{\bk^\prime\lambda^\prime}
(\bx) \mathcal{G}^{(0)}_{\lambda^\prime} (\bk^\prime,i\omega_n) \nonumber\\
&&+ V_0 \mathcal{G}^{(0)}_\lambda (\bk,i\omega_n) 
\psi^\ast_{\bk\lambda} (\bx)
\left(
\sum_{\bq\lambda^{\prime\prime}} V_0 
\psi^\ast_{\bq\lambda^{\prime\prime}} (\bx)
\mathcal{G}^{(0)}_{\lambda^{\prime\prime}} (\bq,i\omega_n)
\psi_{\bq\lambda^{\prime\prime}} (\bx)
\right)
\psi_{\bk^\prime\lambda^\prime} (\bx)
\mathcal{G}^{(0)}_{\lambda^\prime} (\bk^\prime,i\omega_n) \nonumber\\
&&+ \ldots \nonumber\\
&&+ V_0 \mathcal{G}^{(0)}_\lambda (\bk,i\omega_n) 
\psi^\ast_{\bk\lambda} (\bx)
\left(
\sum_{\bq\lambda^{\prime\prime}} V_0 
\psi^\ast_{\bq\lambda^{\prime\prime}} (\bx)
\mathcal{G}^{(0)}_{\lambda^{\prime\prime}} (\bq,i\omega_n)
\psi_{\bq\lambda^{\prime\prime}} (\bx)
\right)^{\ell-1}
\psi_{\bk^\prime\lambda^\prime} (\bx)
\mathcal{G}^{(0)}_{\lambda^\prime} (\bk^\prime,i\omega_n) \nonumber\\
&&+ \ldots ,
\end{eqnarray}
which is recognized as a geometric series, whose sum can be cast in
the form of Eq.~(\ref{eq:DysonT}).
\end{widetext}}

\section{Expansion of the Bloch wavefunctions in the sublattice representation}
\label{app:tightbinding}

At the origin of the $A$ sublattice, $\br=\bz$ say, the Bloch wavefunctions can
be expanded as
\begin{subequations}
\label{eq:psiexpansion}
\begin{eqnarray}
\psi_{\bk A}(\bz) &=& \phi_A^{(0)} + \phi_A^{(1)} \beta_\bk^{(1)} +
\phi_A^{(2)} \beta_\bk^{(2)} + \ldots \\
\psi_{\bk B}(\bz) &=& ~~~~~~~~~ \phi_B^{(1)} \gamma_\bk^{(1)} +
\phi_B^{(2)} \gamma_\bk^{(2)} + \ldots ,
\end{eqnarray}
\end{subequations}
where $\beta_\bk^{(n)}$ and $\gamma_\bk^{(n)}$ are basis functions of the
trivial irreducible representation of the point group $D_{6h}$ and $D_{3h}$,
respectively. In particular, one finds $\gamma_\bk^{(1)} \equiv \gamma_\bk$,
Eq.~(\ref{eq:structfac}), while $\beta_\bk^{(1)} \equiv \beta_\bk$, with
\begin{equation}
\beta_\bk =
{\sum_j}^\prime e^{i\bk\cdot\bR_j},
\label{eq:structfacbeta}
\end{equation}
where the prime restricts the summation to all next nearest neighbors in the
direct lattice, \emph{i.e.} $|\bR_j|=\sqrt{3}a$. 

Because of the rapid decrease of the gaussian pseudoatomic wavefunction,
Eq.~(\ref{eq:gaussian}), one may safely truncate the expansions
Eqs.~(\ref{eq:psiexpansion}) to the first terms, thereby obtaining
\begin{subequations}
\label{eq:psiexpansion1}
\begin{eqnarray}
\psi_{\bk A}(\bz) &\approx& \frac{1}{\sqrt{N}} \phi(\bz) ,\\
\psi_{\bk B}(\bz) &\approx& \frac{1}{\sqrt{N}} \phi(\delta_1) \gamma_\bk .
\end{eqnarray}
\end{subequations}
The latter can be used in the expansion of $\Re G^0 (\bx,\bx,\omega=0)$
appearing in the resonance condition, Eq.~(\ref{eq:rescond}), which for a
site-like impurity reads
\begin{eqnarray}
\Re G^0 (\bx,\bx,0) &=& \frac{s}{t} \sum_{\bk\lambda} |\psi_{\bk\lambda}(\bx)|^2
\nonumber\\
&&+
\frac{1}{t} \sum_\bk \left( \frac{1}{\gamma_\bk} \psi^\ast_{\bk A} (\bx)
\psi_{\bk B} (\bx) + \mathrm{H.c.} \right). \nonumber\\
\label{eq:G0expl}
\end{eqnarray}
Inserting Eqs.~(\ref{eq:psiexpansion}) in Eq.~(\ref{eq:G0expl}) and
Eq.~(\ref{eq:rescond}), one eventually obtains the estimate
Eq.~(\ref{eq:rescond1}) for the impurity strength required to develop a
resonance at $\omega=0$ in the site-like case.

A similar expansion holds in the bond-like and in the hollow-like cases,
Eqs.~(\ref{eq:rescond2}) and (\ref{eq:rescond3}), respectively involving the
constants
\begin{subequations}
\begin{eqnarray}
A_b &=& -\frac{1}{N} \sum_\bk \left( e^{i(\bk\cdot\delta_3 - \theta_\bk)} +
\mathrm{H.c.} \right) \approx 0.67 ,\\
A_h &=& \frac{1}{N} \sum_\bk \left( e^{2i\theta_\bk} \gamma_\bk + \mathrm{H.c.}
\right) \approx 2.35,\\
B_h &=& \frac{1}{N} \sum_\bk |\gamma_\bk|^2 = 3,
\end{eqnarray}
\end{subequations}
\modified{where $e^{i\theta_\bk}$ is defined by Eq.~(\ref{eq:expstructfac}).}

\section{Current density vector within the tight-binding approximation}
\label{app:current}

Here, we summarize some of the results employed to derive the expression of the
conductivity in Sec.~\ref{ssec:conductivity} within the tight binding
approximation outlined in Appendix~\ref{app:tightbinding}. We begin by reminding the
explicit expression of the Fourier transform of the paramagnetic component of
the density current vector in reciprocal space \cite{Bruus:04}
\begin{equation}
\tilde{\bJ}^\nabla (\bk) = -\frac{e}{2m} \int \frac{d\bq}{(2\pi)^2} (2\bq + \bk)
c_\bq^\dag c_{\bk+\bq} .
\end{equation}
In the homogeneous limit ($\bk=\bz$), one has \cite{Paul:03}
\begin{equation}
\tilde{\bJ}^\nabla (0) = \frac{e}{i\hbar} [H,\br] ,
\label{eq:Paul}
\end{equation}
where $H$ is the system's Hamiltonian, including the impurity contribution.
Exploiting Eq.~(\ref{eq:Paul}), one finds
\begin{equation}
\langle \bk\alpha | \tilde{\bJ}^\nabla (0) | \bk^\prime \beta \rangle =
ie\frac{ta}{\hbar}\delta_{\bk^\prime,\bk+\bG} e^{i\bG\cdot\delta_\beta}
\bh_{\alpha\beta} (\bk) ,
\end{equation}
where $\alpha,\beta\in\{ A,B\}$, $\bG$ is a vector of the reciprocal lattice,
and $\delta_\beta=\bz$ if $\beta=A$, and $\delta_\beta=\delta_3$ if $\beta=B$.
Due to the discrete translational invariance and the hermiticity of the current
density operator, the adimensional matrix elements $\bh_{\alpha\beta}(\bk)$
fulfill the additional properties
\begin{subequations}
\begin{eqnarray}
\bh_{AA}(\bk) &=& h_{BB}(\bk),\\
\bh_{AA}(\bk) &=& -h_{AA}(-\bk),\\
\label{eq:hAB}
\bh_{AB}(\bk) &=& -h_{BA}(-\bk).
\end{eqnarray}
\end{subequations}
Moreover, the off-diagonal elements afford the explicit expression
\begin{eqnarray}
\bh_{AB}(\bk) &=& -\frac{i}{a} \nabla_\bk\gamma_\bk \nonumber\\
&=&  \frac{1}{a} \sum_{\ell=1}^3 \delta_\ell e^{i\bk\cdot\delta_\ell} ,
\end{eqnarray}
to leading order in the overlap parameter $s$, where use has been made of
Eq.~(\ref{eq:structfac}),  which, together with Eq.~(\ref{eq:hAB}), yields the
off-diagonal terms of the matrix elements.

In order to find the diagonal terms, it is useful to observe that the
pseudoatomic wavefunctions introduced in Appendix~\ref{app:tightbinding} are
cylindrically symmetric. This implies the following overlap and dipole element
for pseudoatomic wavefunctions centered on nearest neighbor sites
\begin{subequations}
\begin{eqnarray}
\int d\br \,\phi(\br)\,\phi(\br\pm\delta_\ell) &=& s,\\
\label{eq:dipole}
\int d\br \,\phi(\br) \,\br\, \phi(\br\pm\delta_\ell) &=& \mp\frac{1}{2} s\delta_\ell,
\end{eqnarray}
where $s$ is the band asymmetry parameter (Appendix~\ref{app:tightbinding}) and
$\ell=1,2,3$, or, more compactly,
\begin{equation}
\int d\br \,\phi(\br) \,\br \,\phi(\br-\br^\prime) = \frac{1}{2} \br^\prime
\int d\br \,\phi(\br) \,\phi(\br-\br^\prime).
\end{equation}
\end{subequations}
Making use of Eq.~(\ref{eq:dipole}), one eventually finds
\begin{eqnarray}
\bh_{\alpha\alpha} (\bk) &=& -\frac{i}{a} \frac{s}{2}
\nabla_\bk\beta_\bk\nonumber\\
&=& \frac{s}{2a} \sum_{\ell=1}^3 \sum_{m=1,m\neq\ell}^3
(\delta_\ell-\delta_m) e^{i\bk\cdot(\delta_\ell-\delta_m)} ,
\end{eqnarray}
with $\beta_\bk$ given by Eq.~(\ref{eq:structfacbeta}).

\begin{small}
\bibliographystyle{apsrev}
\bibliography{a,b,c,d,e,f,g,h,i,j,k,l,m,n,o,p,q,r,s,t,u,v,w,x,y,z,zzproceedings,Angilella}
\end{small}

\end{document}